\documentclass[oneocolumn,preprintnumbers,amsmath,amssymb,nofootinbib]{revtex4}
\pdfoutput=1

\usepackage{amsmath,latexsym,amssymb,amsfonts}
\usepackage{graphicx,color}
\usepackage{bm}
\usepackage{gensymb}

\begin{document}

\title{Explicit construction of Penrose diagrams for black hole to white hole transition with spacelike thin shells}
\author{
Wei-Chen Lin$^{a,b}$,
Dejan Stojkovic$^{e}$
and
Dong-han Yeom$^{a,c,d,f}$
}
\affiliation{
$^{a}$Center for Cosmological Constant Problem, Extreme Physics Institute, Pusan National University, Busan 46241, Republic of Korea\\
$^{b}$Department of Physics, Pusan National University, Busan 46241, Republic of Korea\\
$^{c}$Department of Physics Education, Pusan National University, Busan 46241, Republic of Korea\\
$^{d}$Research Center for Dielectric and Advanced Matter Physics, Pusan National University, Busan 46241, Republic of Korea \\
$^{e}$HEPCOS, Department of Physics, SUNY at Buffalo, Buffalo, NY 14260, USA  \\
$^{f}$Leung Center for Cosmology and Particle Astrophysics, National Taiwan University, Taipei 10617, Taiwan
}

\begin{abstract}
In this article, we explicitly construct the coordinates associated with the Penrose diagram in spacetimes connected via a spacelike thin shell in the following two examples: the generalized black-to-white hole bounce with mass difference and the Schwarzschild-to-de Sitter transition. We point out the issue of the first junction condition in the Penrose diagram constructed by cutting and pasting analytically known metrics with spherical symmetry by a static spacelike thin shell.  With the goal of a global conformal coordinate chart associated with the corresponding Penrose diagram without discontinuity at the thin shell, we give a procedure consisting of three conformal transformations that serve different purposes. The first two of them are used to generate a continuous coordinate patch covering the entire thin shell, and therefore, the Penrose diagram can be constructed properly by patches with overlapping. The third transformation removes any coordinate singularity reintroduced by the first two transformations at the event horizons.

\end{abstract}

\maketitle

\newpage

\tableofcontents

\section{Introduction}\label{Sec:Intro}

 There are several interesting models in the literature that can be constructed by the cut-and-paste technique of the space-like thin shells. The space-like thin shell region is usually understood as a surface where the quantum gravitational effects might be important. In the absence of a self-consistent theory of quantum gravity, the thin-shell approximation is considered to be a good description of physics right before and after the region where quantum gravitational effects are dominant.

The transition from one Schwarzschild phase to another was discussed in several papers. A cut-and-paste procedure of the space-like slice that covers both inside and outside the horizon was considered in \cite{Ashtekar:2005cj} in the context of the loop quantum gravity. Recently, this was reemphasized by \cite{Haggard:2014rza}, in the so-called black hole fireworks models. The procedure can be described well by using the thin-shell approximation \cite{Brahma:2018cgr}.

A more conservative approach is a black hole to a white hole transition, where one cuts and pastes only the regions inside the event horizon. In this case, we may have several choices. First, one can connect from a black hole to a de Sitter-like phase, and then the de Sitter-like phase is connected to a white hole phase. This is a typical regular black hole model with two (inner and outer) horizons \cite{Frolov:1988vj, Hayward:2005gi}. This model can be derived from loop quantum gravity inspired models \cite{Bojowald:2018xxu}. One can even consider evaporating regular black holes with two horizons using the Vaidya-de Sitter junction of thin-shells \cite{Brahma:2019oal} or numerical simulations \cite{Hwang:2012nn}. Second, one can connect from a black hole phase to a white hole phase directly; hence, there is only one (outer) horizon. This model is also justified from the string \cite{Jusufi:2022uhk} and loop quantum gravity inspired models \cite{Ashtekar:2018lag}, where this is well approximated by thin shells \cite{Bouhmadi-Lopez:2019hpp}.

In this paper, while we do not evaluate the pros and cons of each approach, we admit that the thin-shell technique is a very useful and universal tool to describe quantum gravity inspired ideas. However, at the same time, it is not often emphasized in literature how non-trivial the task is to construct global coordinates that cover both sides of the thin shell. Of course, if the Israel junction conditions are satisfied \cite{Israel:1966rt}, we can be sure that the Einstein equations and energy conservation will be satisfied; hence, locally the description is fine, and one can understand the Penrose diagram piece-wisely (e.g., see \cite{Blau:1986cw}). However, if we do not have a global coordinate chart, it will be difficult to analytically describe the behavior of geodesics that cross the thin shell. However, the dynamics of geodesics that cross the thin shell will be very important to evaluate whether the model is physically consistent or not \cite{Hong:2022thd}.

In this paper, we explore this question by considering two different cut-and-pasted spacetimes with static spacelike junctions: the generalized black-to-white hole bounce \cite{Hong:2022thd} and the Schwarzschild-to-de Sitter transition \cite{ Frolov:1988vj}, in which a thin shell located at some constant areal radius deep inside the Schwarzschild event horizon is assumed. Since we already know the global conformal coordinates of the Schwarzschild solution and the de Sitter space, respectively, instead of starting from constructing a local coordinate chart covering the shell, we start our analysis by utilizing the known Kruskal-like coordinates for the above two spacetimes. Furthermore, the compactified versions of these coordinates, which give the Penrose diagrams of the two spacetimes, will be useful in our analysis. We will comment more on the method adopted in this work and the standard construction of coordinates around the thin shell in Sec.~\ref{Sec:BHtoWH}.  

By constructing from the Kruskal-like coordinates of the spacetimes separated by the shell, we find that two conformal transformations are required to obtain a coordinate chart covering the thin shell unless a special condition is met. The first transformation serves as the removal of the unwanted area, while the second one reconciles the two different coordinate systems on two sides of the shell by removing an implicit discontinuity in the Penrose diagram of this type. However, the second transformation brings a different kind of coordinate singularity back to the event horizon. Depending on the parameters of the spacetimes separated by the thin shell, this coordinate singularity at the event horizon creates a degeneracy such that timelike vectors are either transverse or parallel to the event horizons in the resulting Penrose diagram. We give geometric explanations for the discontinuity of this special type of coordinate singularity and draw a comparison to the Reissner-Nordstr{\"o}m solution. To solve this coordinate singularity reintroduced by the second transformation, we provide a general formalism of the third transformation, which preserves the continuity of coordinates at the junction. By using this formalism, we then give the conformal coordinates in which the metric components are regular at all event horizons, providing that they can be defined by taking the limit of the function there. A 
 cartoon picture summarizing the three transformations is given in Fig.~\ref{fig:Summary}.   

Now, one can ask what the motivation behind this work is. Several models have been proposed to resolve the central singularity of black holes thanks to quantum bounces \cite{Ashtekar:2018lag}, but no one could describe evaporating black holes using these models. If there exists incoming energy flux into the black hole, this will pass the bouncing surface inside the event horizon. Then, what is the back-reaction of Hawking radiation? How can we draw the consistent causal structure or the Penrose diagram of an evaporating black hole? After we know the causal structure, we will check whether these causal structures are consistent or not; we will further ask whether these causal structures can resolve traditional issues such as the information loss paradox. The Vaidya-like approximation might be the first trial to describe the correct causal structure in this context. As long as the mass parameter slowly varies, the Vaidya approximation will be a reasonable attempt to draw the global Penrose diagram. It will be very convenient to do this approximation if we have the global coordinates that cover both sides near the bouncing point. We believe that the present work is the first step toward a consistent description of a quantum-gravity-inspired evaporating black hole model. Especially, the result of the second transformation turns out to give the analytic continuation of the Kruskal-like coordinates through the shell, which nonetheless are only regular at the event horizons of one side of the cut-and-paste spacetime. We also ponder that extending the range of the Kruskal-like coordinates and studying their features facilitate the study of quantum field theory under this type of background spacetime.   

Before entering the organization of this paper, we would like to clarify the following points. Firstly, notice that Israel junction conditions are strictly followed in the construction of the above two cut-and-pasted spacetime models in Ref.~\cite{Hong:2022thd, Frolov:1988vj}. It means that if there exists a continuous coordinate system $x^{\alpha}$ covering a part of spacetime containing the shell, the metric components in the coordinate basis of $x^{\alpha}$ must be at least $\mathcal{C}^{0}$ continuous at the junction. This coordinate-dependent statement of the spacetime metric at the junction is, in fact, equivalent to the coordinate-invariant statement that the induced metric on both sides of the junction must be the same. For a detailed explanation of this point, we refer the readers to Chapters 3.7.1 and 3.7.2 of Ref.~\cite{Poisson:2009pwt}. 

Secondly, we must emphasize that the continuity of the metric component (in a continuous coordinate chart) at the junction mentioned previously, or equivalently the first Israel junction condition, is imposed as a condition directly to the models, and especially it has nothing to do with the energy condition of the shell. In fact, the spacelike shells violate the \textbf{null energy condition} (NEC) in both of the models considered since the equation of state $P/\sigma$ of the shell is negative, where $P$ is the pressure and $\sigma$ is the tension. (The stability analysis of the spacelike thin shell is first introduced in Ref.~\cite{Balbinot:1990zz}, in which the authors discuss the thin shell approximation of the Schwarzschild-to-de Sitter transition. Notice that in Ref.~\cite{Balbinot:1990zz}, the pressure $P$ is called transverse pressure, and the tension $\sigma$ is called axial pressure.)  For further detail, we refer the readers to proposition 1 in Ref.~\cite{Maeda:2023oxl}, and also Section~3.1 therein where the energy condition of a thin shell in the simplest type of black-to-white hole bounce is discussed. Meanwhile, we notice that another way to discuss the energy conditions and junction conditions exists, for instance, Ref.~\cite{Marolf:2005sr}. In Ref.~\cite{Marolf:2005sr}, the junction conditions are not assumed, while in contrast, the existence of a continuous coordinate system $x^{\alpha}$ mentioned above is given to have ansatzes for spacetime metrics. In this context, the authors show that satisfying the NEC at the junction leads to the continuity of the spacetime metric at the same place. However, this was not a demonstration of violation of the NEC leading to discontinuity of the spacetime metric at the thin shell. We also notice that misconception about this issue frequently exists.

Thirdly, one might argue that the Gaussian normal coordinates can always be constructed, which serve as a continuous coordinate system $x^{\alpha}$ covering the part of the spacetime containing the shell, so trajectories or components of any tensor can be faithfully represented in such a coordinate system. We do not disagree with this argument; however, to construct a possible global coordinate system, the relation between such Gaussian normal coordinates and the Schwarzschild coordinates/static coordinates of a de Sitter space still needs to be specified. In fact, for the models considered here, since the shells are always located at some constant $r$ inside the Schwarzschild horizon and the spacetime has spherical symmetry, a Gaussian normal coordinate system respecting the same spherical symmetry should be of the following form $\{x^{0}(r), x^{1}(t), \theta, \phi\}$. However, the further subtlety is that the $\{r, t\}$ coordinates on the two spacetimes separated by the thin shell do not naturally compose a consistent continuous coordinate system across the shell, and they should be labeled as  $\{r_{\pm}, t_{\pm}\}$ to distinguish the two regimes separated by the shell. Then, to have the Gaussian normal coordinates $\{x^{0}, x^{1}, \theta, \phi\}$, even just locally, one still has to obtain the relations $\{x^{0}=A_{+}(r_{+}), x^{1}=B_{+}(t_{+})\}$ and  $\{x^{0}=A_{-}(r_{-}), x^{1}=B_{-}(t_{-})\}$, where $A_{\pm}$ and $B_{\pm}$ are some functions. Later we will see that for the cut-and-paste spacetimes preserving spherical symmetry, the issue is rooted in the relation between $t_{+}$ and $t_{-}$.

The paper is organized as follows: in Sec.~\ref{Sec:BHtoWH} we study a symmetric case, in which the spacetimes on two sides of the shell are both of the Schwarzschild solutions, but their mass parameters can be different. We first show that a simple cut-and-paste procedure creates an implicit discontinuity at the thin shell in the resulting Penrose diagram. We next demonstrate that to fix this issue, one unavoidably reintroduces a special kind of coordinate singularity back to the event horizon. In Sec.~\ref{Sec.BHtodS}, we study an asymmetric case, the Schwarzschild-to-de Sitter transition. While we show that similar problems happen in this type of model, we also discuss the difference between these two models. In Sec.~\ref{Sec:discussions}, we first give the geometric explanations for the discontinuity at the thin shell and the special kind of coordinate singularity created by the transformation fixing the discontinuity. We then draw the comparison to the Reissner-Nordstr{\"o}m solution and its Penrose diagram. In the second half of Sec.~\ref{Sec:discussions}, we give a formalism for the third conformal transformation, which preserves the continuity of the coordinates at the shell. Using this formalism, we show that there is no simple $r$-dependent global conformal coordinate chart for the models considered in this work and then provide a more complicated global conformal coordinate system in which a regular metric can be defined everywhere. We put our conclusion in Sec.~\ref{Sec:conclusion}.

We work in the Planck unit: $G=c=\hbar=1$.

\section{The Penrose diagram of a Generalized Black-to-White Hole Bouncing Model}\label{Sec:BHtoWH}

In the generalized black-to-white hole bouncing model,  an effective thin shell connects a black hole phase to a white hole phase deep inside the event horizon. On each side of the shell, the metric of spacetime is given by the usual Schwarzschild metric, but the mass parameter can be different. The stability issue of this type of model was studied in Ref. \cite{Hong:2022thd}, and a static thin shell was found to be possible if there is no restriction on the equation of state of the thin shell.

Now, we are looking for a global conformal coordinate system for this cut-and-pasted spacetime, or at least a local coordinate system covering the entire thin shell. A standard method might be described as follows. First, we construct the local Gaussian coordinates described in the previous section. After this, we try to find a suitable coordinate transformation that makes the new coordinate system regular at all event horizons in this spacetime if such a coordinate system exists. This global coordinate system should be very complicated if it were found. Thus, the relation between this global coordinate system and the Kruskal-Szekeres coordinates might be unclear. Given these issues and the fact that we already have the well-known global coordinates for the Schwarzschild spacetime, we take a different strategy by reversing the above mentioned procedure. That is, we start directly by using the Kruskal-Szekeres coordinates of the two Schwarzschild spacetimes separated by the thin shell and then perform modification to enforce continuity of the metric components at the shell, which in turn manifests the first Israel junction condition as mentioned in the first point in Sec.~\ref{Sec:Intro}. Meanwhile, we also find that the Penrose diagram is particularly useful in our construction due to the following reasons. Firstly, since the Kruskal-Szekeres coordinates have the range $(-\infty, \infty)$, a compactified version of them is required when combining multiple Kruskal-Szekeres coordinate systems. For the Schwarzschild solution (and later the de Sitter space), the Penrose diagram is simply created by taking the inverse tangent function of the null Kruskal-Szekeres coordinates, which naturally fulfills this condition. Later we will see that the Penrose diagram can lead us to the required transformations and offers geometrical interpretations of the transformations performed. 
(Generally speaking, determining a Penrose diagram for a general spacetime does not require knowledge of a global coordinate system on that spacetime. We thank one of the anonymous referees for emphasizing this point to us.)

Before we discuss the construction of the Penrose diagram of the generalized black-to-white hole bouncing model, we quickly review the typical construction of the Penrose diagram of the maximally extended Schwarzschild solution as a way to set up the notations and conventions.

\subsection{The maximally extended Schwarzschild solution}

In spherical coordinates $\{t, r, \theta, \phi \}$, the Schwarzschild metric is given by
\begin{equation}\label{Schwarzschild_metric}
d s^2=-f(r)d t^2+f^{-1}(r)d r^2+r^2d\Omega^2, 
\end{equation}
where $f(r)=1-2M/r$ and $d\Omega^2=d\theta^2 +\sin^2{\theta}d\phi^2$.  Inside the event horizon $r<2M$, this form of metric can be interpreted as the metric of the \textbf{Kantowski-Sachs} spacetime such that $t$ becomes a spatial coordinate while $r$ is the temporal coordinate. Thus besides at the event horizon, which is a coordinate singularity of the metric (\ref{Schwarzschild_metric}), in the coordinate systems $\{t, r, \theta, \phi \}$
the four-velocity of an observer moving along with a radial timelike geodesic is given by
\begin{equation}\label{radial_geodesic_Deq}
\mathcal{U}^{\alpha}=\left(\frac{d t}{d \tau}, \frac{d r}{d \tau}, \frac{d \theta}{d \tau}, \frac{d \phi}{d \tau} \right)=\left(\frac{E}{f(r)}, -\epsilon \sqrt{E^2-f}, 0, 0\right),
\end{equation}
where $\tau$  and $E$ are the proper time and the conserved energy per unit mass associated with the geodesic, respectively. Also notice that we choose $\epsilon=1$ for the infalling geodesics and $\epsilon=-1$ for the outgoing geodesics.  For $E^2 \geq 1$, the conserved quantity $E$ can be related to the observer's velocity at infinity $v_{\infty}$ as $E^2=1/(1-v^2_{\infty})$ by noticing that the Schwarzschild metric (\ref{Schwarzschild_metric}) reduces to the Minkowski metric at $r \rightarrow \infty$. For $E \leq 1$, $E$ is related to the maximal radius $R$ the geodesic can reach by the relation $E^2=1-2M/R$. Meanwhile, when $E=0$, the four-velocity (\ref{radial_geodesic_Deq}) has real components only inside the event horizon, which corresponds to a special group of radial geodesics comoving with respect to the Kantowski-Sachs spacetime. In particular, the trajectories of those special geodesics can be labeled by constant $t$. We will utilize this group of geodesics later when constructing the Penrose diagram of the black-to-white hole bouncing model.

\begin{figure}[h!]
	\centering
	\includegraphics[scale=0.4]{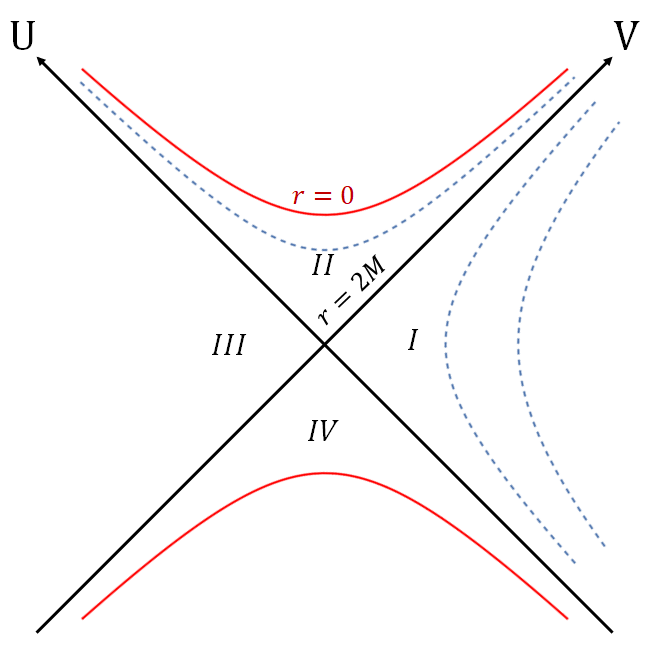}
	\caption{A Kruskal diagram. When considering only the infalling geodesics, the relevant regions are regions $I$ and $II$. }  
	\label{fig:Simple_KS}
\end{figure}

Next, the maximally extended Schwarzschild solution without coordinate singularity at the event horizon is given by the null \textbf{Kruskal-Szekeres} (KS) coordinates $(V, U)$ through a series of coordinate transformations \cite{Carroll:2004st}. (Whenever the acronym KS is used, we refer to Kruskal-Szekeres (coordinates) instead of the Kantowski-Sachs (spacetime).) In terms of the Schwarzschild coordinates $(t,r)$, they are given by 
\begin{equation}\label{Null_KS_in_r_t}
V=\pm e^{(t+r^{*})/4M};    \quad U=\mp e^{-(t-r^{*})/4M}
\end{equation}
where $r^{*}$ is the \textbf{tortoise coordinate}
\begin{equation}
r^{*} \equiv r + 2M \ln \left|\frac{r}{2M}-1\right|,
\end{equation}
and the plus/minus signs are directly determined by the region considered in a Kruskal diagram. See Fig.~\ref{fig:Simple_KS}. From Eq.~(\ref{Null_KS_in_r_t}), one can show that the null coordinates $(V, U) $ satisfy the relations
\begin{equation}\label{Kruskal_UV(r)}
UV=\left(1-\frac{r}{2M}\right)e^{r/2M},
\end{equation}
and 
\begin{equation}\label{Kruskal_U/V}
\frac{U}{V}=\mp e^{-t/2M},
\end{equation}
where the minus sign refers to the $r>2M$ region, and the plus sign refers to the $r<2M$ region.
The metric of a maximally extended Schwarzschild solution in terms of the null Kruskal coordinates $(V, U) $ is given by 
\begin{equation}\label{Kruskal_in_U_V}
d s^2=-\frac{16M^3}{r}e^{-r/2M}(d V d U+d U d V)+r^2 d \Omega^2,
\end{equation}
which is well-defined everywhere except only at the singularity. Lastly, to obtain the Penrose diagram of the Schwarzschild solution, one simply converts the null KS coordinates to the compactified coordinates of the Penrose diagram $(\tilde{V}, \tilde{U})$ through the transformation:
\begin{equation}\label{KS_to_Penrose}
(\tilde{V}, \tilde{U})=\left(\tan^{-1}V, \tan^{-1}U \right).  
\end{equation}
In the rest of this work, we broadly name this type of coordinates, here $(\tilde{V}, \tilde{U})$ for the standard Schwarzschild solution, as ``the coordinates associated with the Penrose diagram'' in the sense that these coordinates are constructed step by step to form a compactified diagram.

\subsection{The generalized black-to-white hole bouncing model}\label{subSec:BHtoWH_model}

As mentioned earlier at the beginning of this section, a way to generalize the black-to-white hole bouncing model is using the thin shell approximation. On each side of the shell, the metric ansatz is given by 
\begin{equation}\label{Schwarzschild_metric_inside_BWH}
d s_{\pm}^2=-(-f_{\pm})^{-1}d r_{\pm}^2+(-f_{\pm})d t_{\pm}^2+r_{\pm}^2d\Omega^2, 
\end{equation}
which describes the metric inside the horizon, with
\begin{equation}\label{f(r)_function}
f_{\pm} = 1-\frac{2M_{\pm}}{r},
\end{equation}
where $-$ denotes the black hole phase and $+$ denotes the white hole phase, and the mass parameters generally are not the same: $M_{-} \neq M_{+}$. Notice Eq.~(\ref{Schwarzschild_metric_inside_BWH}) is, in fact, the usual Schwarzschild metric (\ref{Schwarzschild_metric}) but expressed in the form emphasizing the changing the roles of the coordinates $\{t, r\}$ mentioned earlier. 
Under the thin-shell formalism, it has been shown that by fine-tuning the equation of state of the shell, a spacelike thin shell connecting a black hole phase to a white hole phase can be stationary \cite{Hong:2022thd}.

In particular, if we consider only the static case, then from the first junction condition that the induced metric on both sides of the thin shell must be the same, we have the following relation
\begin{equation}\label{induced_metric_relation_BWH}
(-f_{-})d t_{-}^2+r_{-}^2d\Omega^2=(-f_{+})d t_{+}^2+r_{+}^2d\Omega^2. 
\end{equation}
Thus, in an effective model that a black hole phase transitions to a white hole phase at some minimal radius $b$, we must have
\begin{equation}\label{1JC_BWH_r=b}
r_{+}=r_{-}=b.
\end{equation}
and 
\begin{equation}\label{1JC_BWH_t_component}
\sqrt{-f_{-}(b)}d t_{-}=\sqrt{-f_{+}(b)}d t_{+},  
\end{equation}
which further leads to 
\begin{equation}\label{t_relation_BWH}
t_{+}=\sqrt{\frac{f_{-}(b)}{f_{+}(b)}}t_{-},
\end{equation}
where the constant of integration is set to zero. In Ref. \cite{Hong:2022thd}, the effect of Eq.~(\ref{1JC_BWH_t_component}) on timelike radial geodesics was studied and it was found that when $\sqrt{-f_{-}(b)} \neq \sqrt{-f_{+}(b)}$, a timelike radial geodesic loses or gains energy after crossing the thin shell. To be more specific, to have the timelike radial geodesic smoothly crossing the thin shell, the energy parameters $E_{\pm}$ related to it on two sides of the shell satisfy the following energy shifting relation
\begin{equation}\label{Blue_shift_BWH}
E_{+}=\sqrt{\frac{f_{+}(b)}{f_{-}(b)}}E_{-}. 
\end{equation}
Thus the trivial case is when $dt_{-}=dt_{+}$, in which there is no energy shifting for timelike radial geodesics crossing the thin shell, $E_{+}=E_{-}$.    

Here, since our goal is to make a continuous coordinate chart covering the entire thin shell, we also have to study the consequence of Eq.~(\ref{t_relation_BWH}). One can understand the physical meaning of
Eq.~(\ref{t_relation_BWH}) as follows. We first align the origins of the two spatial coordinates $t_{-}$ and $t_{+}$ on each side of the shell. This action corresponds to setting the constant of integration to zero. Next, supposing that we put a ruler on this static shell along the $t-$direction with one end at $t_{-}=t_{+}=0$, then Eq.~(\ref{t_relation_BWH}) gives us the relation of the position of the other end of the ruler measured in $t_{-}$ and $t_{+}$ coordinates. That is, if we measure the length of such a ruler to be $t_{-}=t_{a}$ in coordinate $t_{-}$, then its measurement in coordinate $t_{+}$
must be $t_{+}=\sqrt{f_{-}(b)/f_{+}(b)} \, t_{a}$. 
Therefore, after we align $t_{-}=t_{+}=0$ as the reference point on the static shell, for any point on the shell, its coordinates in the two coordinate systems satisfy Eq.~(\ref{t_relation_BWH}).   

Now, since Eq.~(\ref{Schwarzschild_metric_inside_BWH}) has the same form of the Schwarzschild metric (\ref{Schwarzschild_metric}), to obtain the Penrose diagram of the generalized black-to-white hole bouncing model, we first convert Eq.~(\ref{Schwarzschild_metric_inside_BWH}) to the null KS coordinates to get rid of the coordinate singularities at the event horizons and to have the maximally extended solution of the corresponding spacetime on each side. However, since the two coordinate systems are connected at $r_{+}=r_{-}=b$, a ``good'' Penrose diagram of the above-mentioned model cannot be constructed simply by the transformation (\ref{KS_to_Penrose}) from the null KS coordinates.  In the rest of this section, we show how to construct the corresponding Penrose diagram through a series of coordinate transformations such that the resulting coordinates of the Penrose diagram are continuous around the thin shell. To be more precise, we demonstrate that not only a transformation removing the unwanted $r_{\pm}<b$ region is required, but a second conformal transformation is needed to have the coordinates associated with the Penrose diagram consistent with the first junction condition and thus, form a continuous chart covering the entire thin shell.

\subsection{Removing $r_{\pm}<b$ region}\label{subSec: 1st_conformal_trans}

\begin{figure}[h!]
	\centering
	\includegraphics[scale=0.3]{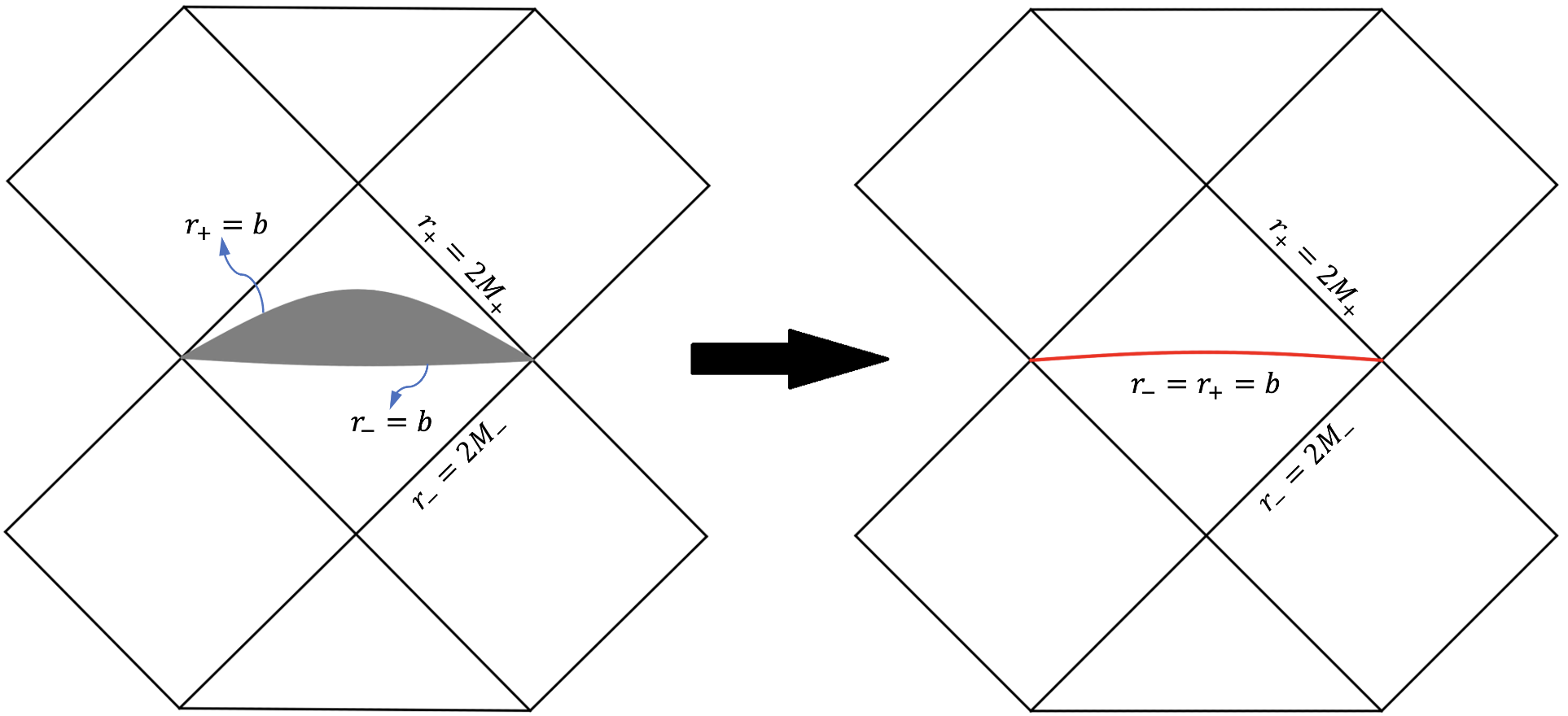}
	\caption{The simple cut-and-paste procedure creating the Penrose diagram of the generalized black-to-white hole bouncing model. Notice that in the resulting Penrose diagram on the right, the static thin shell must be represented by one single spacelike curve. }  
	\label{fig:intro}
\end{figure}

In this part, we first consider a minimal procedure one has to perform to remove the $r_{\pm}<b$ region in the Penrose diagram as shown in Fig.~\ref{fig:intro}. 
It consists of a rescaling transformation, a cutting procedure, and a pasting procedure. We will refer to this type of procedure as the simple cut-and-paste procedure in the rest of this paper.

To begin with, from Eq.~(\ref{Kruskal_UV(r)}), the null Kruskal coordinates $(V_{\pm}, U_{\pm}) $ on each side satisfy the following relation at the transition surface $r_{+}=r_{-}=b$:
\begin{equation}\label{UV_mini_r}
U_{\pm}V_{\pm}=\left(1-\frac{b}{2M_{\pm}}\right)e^{b/2M_{\pm}} \equiv X^2_{\pm},
\end{equation}
where $X_{\pm}=\sqrt{1-b/2M_{\pm}}e^{b/4M_{\pm}}>0$. To remove the region of $r_{\pm}< b$ in the corresponding Penrose diagram, we introduce the following rescaled null Kruskal coordinates
\begin{equation}\label{ModKS}
(V'_{\pm}, U'_{\pm})\equiv \left(\frac{V_{\pm} }{X_{\pm}},  \frac{U_{\pm} }{X_{\pm}}\right).  
\end{equation}
Notice that this transformation is of the form 
\begin{equation}\label{preserving_45_trans}
(V'_{\pm}, U'_{\pm})=\left(g_{\pm}(V), \, h_{\pm}(U)\right),
\end{equation}
where $g_{\pm}(V)$ and $ h_{\pm}(U)$ are some well-behaved functions, for instance, the inverse functions of them exist. Therefore the radial null trajectories are still at $45 \degree$ on the corresponding modified Kruskal diagrams. (In this work, we refer to the transformation keeping radial null trajectories at $45 \degree$ as a conformal transformation.) Moreover, the ratio between the two null Kruskal coordinates is still the same after the rescaling transformation (\ref{ModKS}): 
\begin{equation}\label{U/V_mini}
\frac{U'_{\pm}}{V'_{\pm}}=\frac{U_{\pm}}{V_{\pm}}= \pm e^{-t_{\pm}/2M_{\pm}}.
\end{equation}
Recalling that inside the event horizon, a trajectory with constant $t$ belongs to the special family of geodesics comoving with the interior Kantowski-Sachs spacetime; thus, for the interior part, the rescaling transformation acts as stretching the spacetime manifold along these special geodesics. 

\begin{figure}[h!]
	\centering
	\includegraphics[scale=0.3]{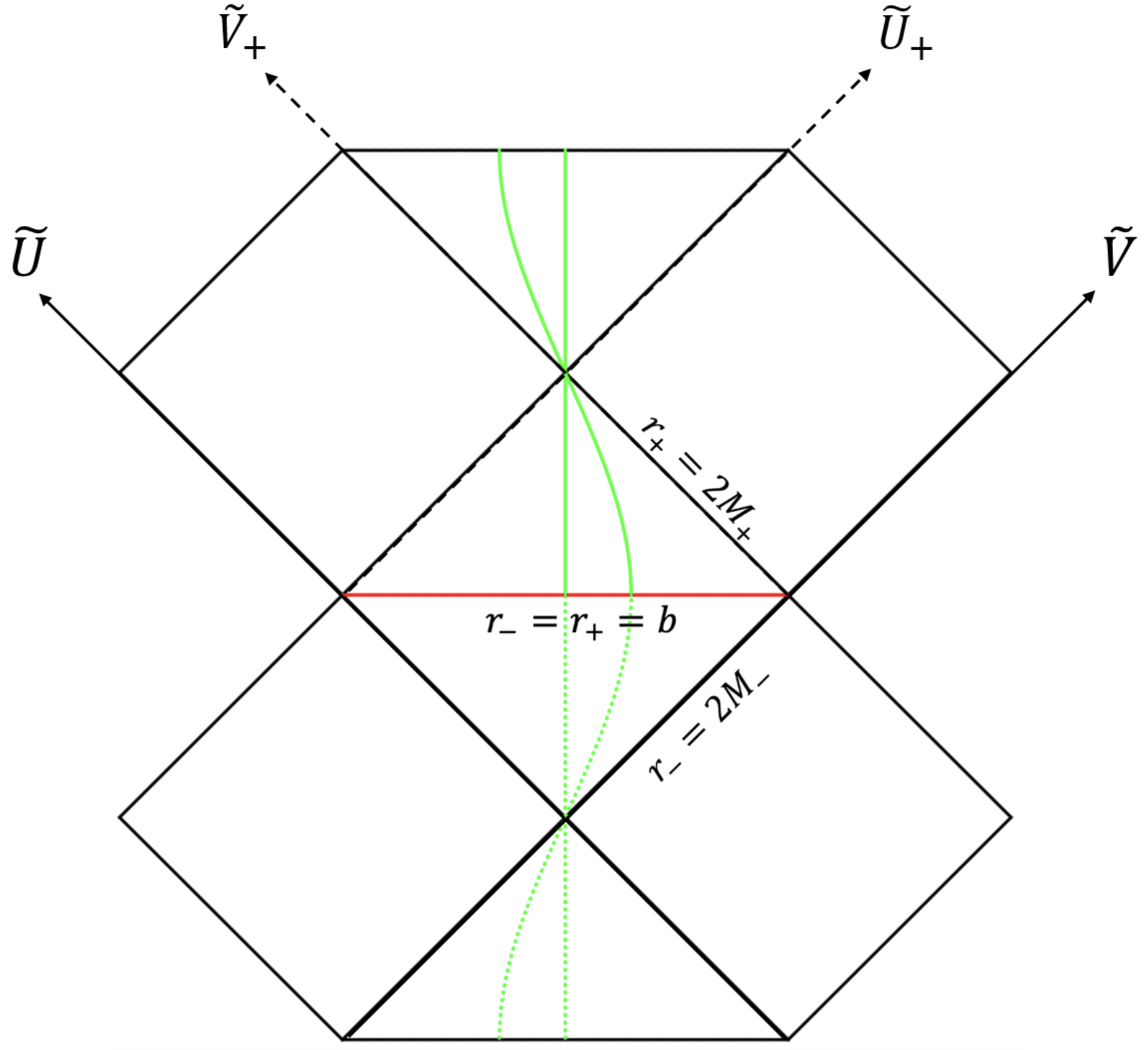}
	\caption{The Penrose diagram with $r_{\pm}<b$ removed by using the simple cut-and-paste procedure.  The seemingly connected green trajectories, which correspond to the special geodesics with constant $t_{\pm}$, are in fact different geodesics except the reference one with $t_{\pm}=0$. Therefore, there is an implicit discontinuity at the thin shell.}  
	\label{fig:first_trans_result}
\end{figure}

Next, to perform the cutting and pasting, we compactify the modified Kruskal diagram by the usual inverse tangent transformation 
\begin{equation}\label{ModKS_to_ModPenrose}
(\tilde{V}_{\pm}, \tilde{U}_{\pm})=\left(\tan^{-1}V'_{\pm}, \tan^{-1}U'_{\pm}\right).
\end{equation}
Due to this rescaling transformation (\ref{ModKS}), the thin shell connecting the black hole phase to the white hole phase has the coordinates $\tilde{V}_{-}+\tilde{U}_{-}=\pi/2$ with $0 \leq \tilde{V}_{-} \leq \pi/2$ from the black hole side, and $\tilde{V}_{+}+\tilde{U}_{+}=-\pi/2$ with $-\pi/2 \leq \tilde{V}_{+} \leq 0$ from the white hole side. Therefore, we can cut out the region with $\tilde{V}_{-}+\tilde{U}_{-}>\pi/2$ and $\tilde{V}_{+}+\tilde{U}_{+}<-\pi/2$, which corresponds to $r_{-}<b$ and $r_{+}<b$ respectively, and then paste the two Penrose diagrams together by following identifications 
\begin{equation}\label{BH_WH_connection_rule_0}
(\tilde{V}, \tilde{U})  \equiv (\tilde{V}_{-}, \tilde{U}_{-}),  
\end{equation}
and
\begin{equation}\label{BH_WH_connection_rule}
(\tilde{V}, \tilde{U}) =
     \left(\tilde{U}_{+}+\frac{\pi}{2}, \tilde{V}_{+}+\frac{\pi}{2}\right).    
\end{equation}
Notice that the relation Eq.~(\ref{BH_WH_connection_rule}) is required to have a consistent $t$ increasing direction. The resulting Penrose diagram of this simple cut-and-paste procedure is given in Fig.~\ref{fig:first_trans_result}. (Notice that our procedure is dissected into three transformations. Here is the intermediate result of the first transformation, which requires further transformation.)

Now, we examine if the coordinates associated with the resulting Penrose diagram have any implicit illness around the thin shell by using the trajectories of the special radial geodesics with constant $t_{\pm}$ mentioned earlier. ( Notice that we will also call ``the coordinates associated with the resulting Penrose diagram'' to be ``the resulting Penrose diagram'' in the rest of the paper for simplicity. )  
To be specific, we choose two segments of those geodesics labeled by $t_{-}=t_{a}$ and $t_{+}=t_{b}$ on each side of the shell, respectively. And we ask what the relation between $t_{a}$ and $t_{b}$ should be for them to form one continuous trajectory on this resulting Penrose diagram. If this relation turns out to be different from Eq.~(\ref{t_relation_BWH}) indicated by the first junction condition, then this resulting Penrose diagram has an implicit discontinuity at the thin shell, such that an apparently continuous trajectory crossing the thin shell on this Penrose diagram is in fact segments of two different trajectories end and start from the thin shell at different locations. In the following, we demonstrate that it is indeed such a case. 

First, in coordinates $(\tilde{V}_{-}, \tilde{U}_{-})$, the trajectory of the geodesics labeled by  $t_{-}=t_{a}$ is given by
\begin{equation}\label{KS_Ober_in_minus}
 \tilde{U}_{-}=\tan^{-1}\left[e^{-t_{a}/2M_{-}} \tan\tilde{V}_{-} \right]. 
\end{equation}
Together with the equation of the thin shell  $\tilde{V}_{-} + \tilde{U}_{-}=\frac{\pi}{2}$ and Eq.~(\ref{BH_WH_connection_rule_0}), we have the coordinates where this particular radial geodesic crosses the thin shell as: 
\begin{equation}\label{KS_crossing_from_minus}
(\tilde{V}, \tilde{U})=(\tilde{V}_{-}, \tilde{U}_{-}) = \left(\tan^{-1}\left[e^{t_{a}/4M_{-}}\right], \tan^{-1}\left[e^{-t_{a}/4M_{-}}\right]\right).
\end{equation}
Next, the trajectory of the geodesics labeled by  $t_{+}=t_{b}$ in coordinates $(\tilde{V}_{+}, \tilde{U}_{+})$ is given by 
\begin{equation}\label{KS_Ober_in_plus}
 \tilde{U}_{+}=\tan^{-1}\left[e^{-t_{b}/2M_{+}} \tan\tilde{V}_{+} \right]. 
\end{equation} 
By noticing that in coordinates $(\tilde{V}_{+}, \tilde{U}_{+})$ the equation of the same thin shell  is now given by $\tilde{V}_{+} + \tilde{U}_{+}=\frac{-\pi}{2}$, we have the coordinates where the geodesics labeled by $t_{+}=t_{b}$ crosses the thin shell as
\begin{equation}\label{KS_crossing_from_plus}
(\tilde{V}, \tilde{U})=\left(\tilde{U}_{+}+\frac{\pi}{2}, \tilde{V}_{+}+\frac{\pi}{2}\right) = ( \tan^{-1}\left[-e^{-t_{b}/4M_{+}}\right]+\frac{\pi}{2}, \tan^{-1}\left[-e^{t_{b}/4M_{+}}\right]+\frac{\pi}{2}), 
\end{equation}
where Eq.~(\ref{BH_WH_connection_rule}) is used.
To have the two segments, Eqs.~(\ref{KS_Ober_in_minus}) and (\ref{KS_Ober_in_plus}), forming one single trajectory on the Penrose diagram with $r_{\pm}<b$ removed as shown in Fig.~\ref{fig:first_trans_result},    
 Eqs.~(\ref{KS_crossing_from_minus}) and (\ref{KS_crossing_from_plus}) must be the same point in the coordinates $(\tilde{V}, \tilde{U})$. Then, by using $\tan^{-1}(-x)=-\tan^{-1}x$ and the following identity to simplify Eq.~(\ref{KS_crossing_from_plus}),  
\begin{equation}\label{Arctan_identity}
\tan^{-1}\frac{1}{x}+\tan^{-1}x=\frac{\pi}{2} \quad if  \ x>0, 
\end{equation}
the aforementioned condition leads to the following relation
\begin{equation}\label{continuity_condition}
\frac{t_a}{M_{-}}=\frac{t_b}{M_{+}},
\end{equation}
which is in contradiction to the relation given by the first junction condition, Eq.~(\ref{t_relation_BWH}), unless $M_{+}=M_{-}$. 
This contradiction means that in the general cases $M_{+} \neq M_{-}$, there still is an implicit discontinuity at the thin shell in the Penrose diagram with $r_{\pm}<b$ removed by the procedure introduced. Thus, we need to use the other coordinate transformation to enforce the relation indicated by the first junction condition, Eq.~(\ref{t_relation_BWH}), when constructing the Penrose diagram of the generalized black-to-white hole bouncing model.  

Lastly, it is worth mentioning that Eq.~(\ref{ModKS}), in fact, contains two different transformations acting separately on each side of the shell. One might wonder if this is the reason for the inconsistency. To get rid of this potential issue, one can use the following transformation to accomplish the same goal, \textit{i.e.} removing $r_{\pm} < b$ region:  \begin{equation}\label{single_rescaling_trans}
(V'_{\pm}, U'_{\pm})\equiv \left(\frac{V_{\pm} }{X},  \frac{U_{\pm} }{X}\right), 
\end{equation}
where $X=\sqrt{X_{+}X_{-}}$ and $X_{\pm}$ are defined in Eq.~(\ref{UV_mini_r}). Notice that if we use Eq.~(\ref{single_rescaling_trans}) instead of Eq.~(\ref{ModKS}), the thin shell will bend to the side having a smaller mass.  Performing a similar calculation, we will still arrive at the same relation Eq.~(\ref{continuity_condition}), so the contradiction is not caused by using Eq.~(\ref{ModKS}). The reason why we choose to use Eq.~(\ref{ModKS}) instead of the more succinct transformation Eq.~(\ref{single_rescaling_trans}) is due to the fact that the second transformation introduced can only work after Eq.~(\ref{ModKS}). See Sec. \ref{Subsec: geometric interpretation} for further discussion.

\subsection{The second transformation fixing the discontinuity at the thin shell}\label{subSec:second_trans_BHtoWH}

\begin{figure}[h!]
	\centering
	\includegraphics[scale=0.3]{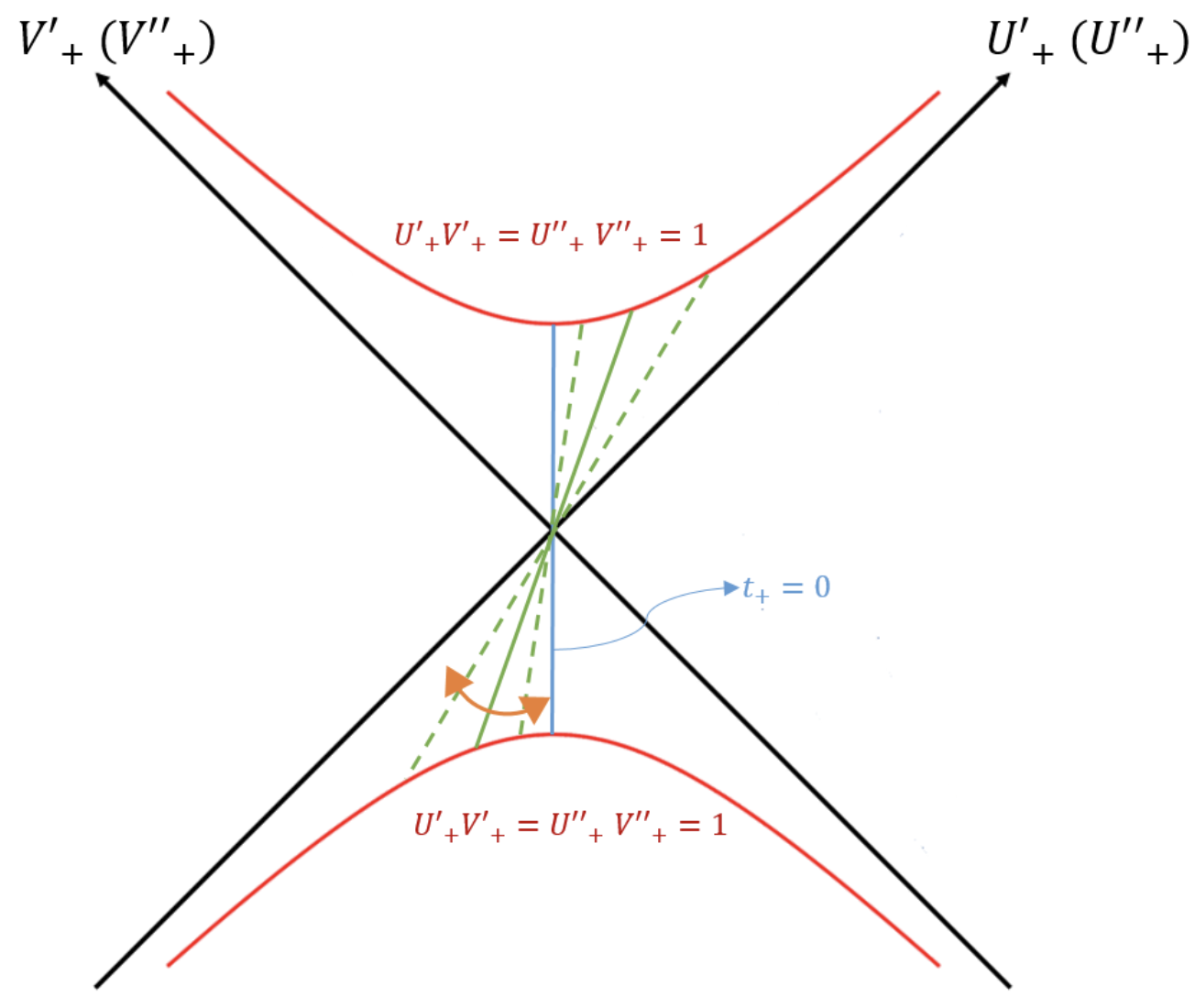}
	\caption{  The effect of a conformal transformation with the form $\left(V''_{+}, U''_{+}\right)=\left(V'_{+} \left| V'_{+} \right| ^{k-1}, U'_{+} \left| U'_{+} \right| ^{k-1} \right)$ with $k \neq 1$. This transformation rotates trajectories with constant $t_{+}$, but keeps the ``locations" of event horizons, the thin shell, and the trajectory of $t_{+}=0$ unchanged.}  
	\label{fig:second_conformal}
\end{figure}

\begin{figure}[h!]
	\centering
	\includegraphics[scale=0.7]{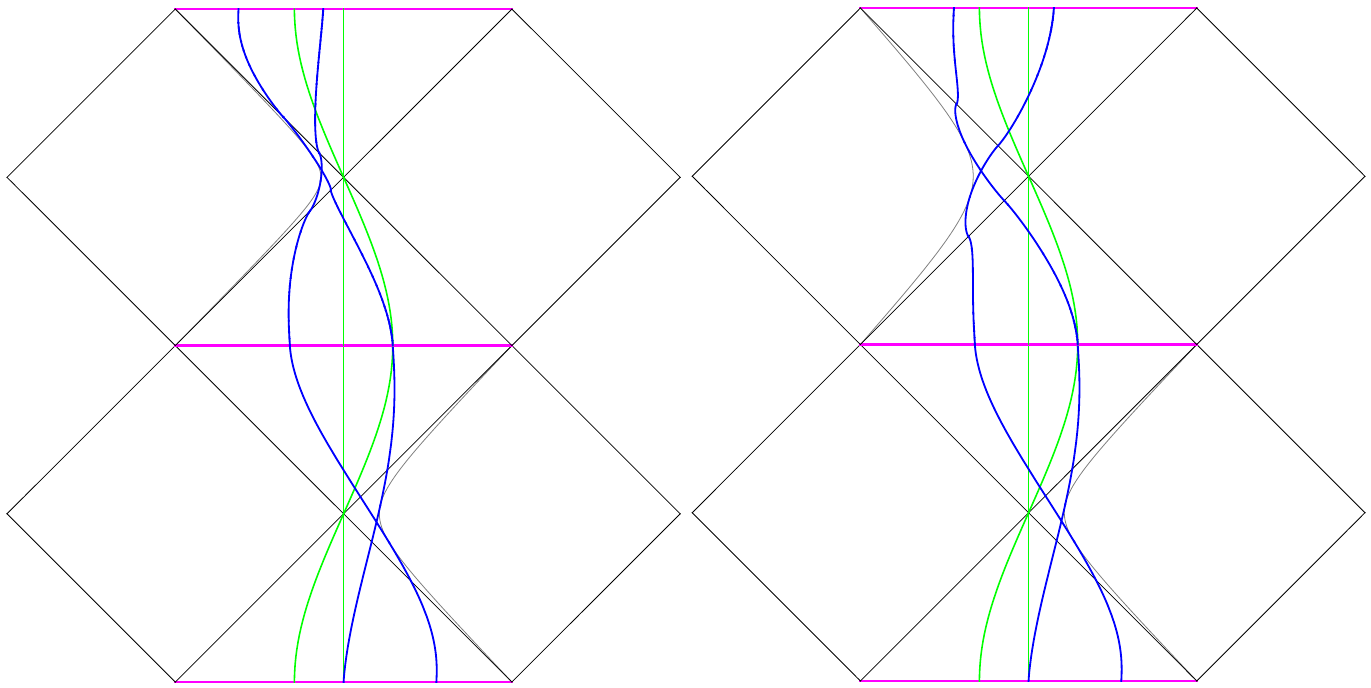}
	\caption{The trajectories of the bounded radial geodesics in Penrose diagrams without discontinuity at the thin shell. The green lines are the special group of radial geodesics with $E=0$, while the blue lines are some radial geodesics specified by some maximal radius (in gray) they can reach. The condition that any timelike radial geodesics should cross the thin shell smoothly is manifested by their trajectories in this resulting diagram as they satisfy the relation (\ref{Smoothly_crossing_condition}) at the thin shell.
 The diagram on the left is a mass-increasing scenario with $M_{+}/M_{-}=1.2$, while the diagram on the right is a mass-decreasing scenario with $M_{+}/M_{-}=0.8$. Notice that the bottom halves of the two diagrams are the same, but the second halves on the top of them are different. In particular, there is a distortion in the trajectories of the general radial geodesics (blue) around the event horizon due to a new type of coordinate singularity generated by the second conformal transformation (\ref{Second_conformal_BHtoWH}). The distortion can be distinguished into the stretching type (the left figure) and the squeezing type (the right figure) from the geometric point of view. See Sec. \ref{Subsec: geometric interpretation} for further discussion.}  
	\label{fig:second_trans_BHtoWH}
\end{figure}

To fix the implicit discontinuity, we can perform the following second coordinate transformation on $(V'_{+}, U'_{+})$ after the first transformation (\ref{ModKS}):
\begin{equation}\label{Second_conformal_BHtoWH}
\left(V''_{+}, U''_{+}\right)=\left(V'_{+} \left| V'_{+} \right| ^{k-1}, U'_{+} \left| U'_{+} \right| ^{k-1} \right),
\end{equation}
where $k \equiv \frac{M_{+}}{M_{-}}\sqrt{f_{+}(b)/f_{-}(b)}$
such that 
\begin{equation}\label{second_conformal_moving_t}
\frac{U''_{+}}{V''_{+}}=\pm e^{-\frac{t_{+}}{2M_{-}}\sqrt{f_{+}(b)/f_{-}(b)}}. 
\end{equation}
This transformation is conformal since constant $V'_{+}$ and $U'_{+}$ lines map to constant $V''_{+}$ and $U''_{+}$ lines respectively, \textit{i.e.} the null geodesics are still at $45\degree$. Furthermore, the ``locations" of event horizons and the thin shell are unchanged in the modified KS diagram. See Fig.~\ref{fig:second_conformal}. From Eq.~(\ref{second_conformal_moving_t}) we can see that this transformation rotates trajectories with constant $t_{+}$ in Fig.~\ref{fig:second_conformal}, but keeps the reference ``position'' $t_{+}=0$ unchanged. Therefore, following the same procedure introduced in the previous subsection with the extra transformation (\ref{Second_conformal_BHtoWH}), a Penrose diagram of the generalized black-to-white hole bouncing model can be constructed without discontinuity around the thin shell. An example of the resulting Penrose diagram with trajectories of general bounded radial geodesics is given in Fig.~\ref{fig:second_trans_BHtoWH}. 

To see the continuity of the metric components in the coordinates of the resulting Penrose diagram $(\tilde{V}, \tilde{U})$, we first rewrite the metric component in terms of $(\tilde{V}, \tilde{U})$ as
\begin{equation}\label{Penrose_metric_lower}
ds^2=-\frac{32M_{-}^3}{r_{-}}e^{-r_{-}/2M_{-}}X_{-}^{2} \left(\frac{1}{\cos^2\tilde{V}\cos^2\tilde{U}}\right)d\tilde{V} d \tilde{U}+r_{-}^2 d \Omega^2
\end{equation}
when $(\tilde{V}, \tilde{U})$ satisfying the relations: $\frac{\pi}{2} \geq \tilde{V} \geq -\frac{\pi}{2}$,  $\frac{\pi}{2} \geq \tilde{U} \geq -\frac{\pi}{2}$ and $\frac{\pi}{2} \geq \tilde{V}+\tilde{U} \geq -\frac{\pi}{2}$, \textit{i.e.} $(\tilde{V}, \tilde{U})$ is a spacetime event of the lower half of the resulting Penrose diagram, Fig.~\ref{fig:second_trans_BHtoWH}. While for the upper half of the resulting Penrose diagram, in which  relations $\pi \geq \tilde{V} \geq 0$,  $\pi \geq \tilde{U} \geq 0$ and $\frac{3\pi}{2} \geq \tilde{V}+\tilde{U} \geq \frac{\pi}{2}$ are satisfied, the metric is given by 
\begin{equation}\label{Penrose_metric_upper}
 ds^2=
 -\frac{32M_{+}^3}{r_{+}}e^{-r_{+}/2M_{+}}\frac{X_{+}^{2k}}{k^2} \left|\left(1-\frac{r_{+}}{2M_{+}}\right)e^{r_{+}/2M_{+}}\right|^{1-k}\left(\frac{1}{\sin^2\tilde{V}\sin^2\tilde{U}}\right) d\tilde{V} d \tilde{U}+r_{+}^2 d \Omega^2, 
\end{equation}
where Eq.~(\ref{metric_after_2nd_trans_in_r}) is used.
Notice that the explicitly $\tilde{V}$ or $\tilde{U}$ dependent factors in Eqs.~(\ref{Penrose_metric_lower}) and (\ref{Penrose_metric_upper}) are only divergent at the infinities as they should be. Then, one can check that Eqs.~(\ref{Penrose_metric_lower}) and (\ref{Penrose_metric_upper}) reduce to the same form at the transition surface $r_{-}=r_{+}=b$, which is the segment $\tilde{V}+\tilde{U}=\frac{\pi}{2}$ in the resulting Penrose diagram. This result shows that the coordinates of the resulting Penrose diagram $(\tilde{V}, \tilde{U})$ indeed form a continuous coordinate chart across the thin shell.

The condition that a timelike radial geodesic must cross the thin shell smoothly is also manifested by its trajectory in the resulting Penrose diagram. This result serves as another confirmation of the continuity of coordinates $(\tilde{V}, \tilde{U})$ around the shell. To be specific, by using the coordinates of the resulting Penrose diagram,  $(\tilde{V}, \tilde{U})$, such a condition can be formulated as
\begin{equation}\label{Smoothly_crossing_condition}
\frac{d\tilde{U}}{d\tilde{V}} \Big|_{r_{-} \to b}=\frac{d\tilde{U}}{d\tilde{V}} \Big|_{r_{+} \to b}.
\end{equation}     
We leave the proof in the companion work \cite{Lin:2023sxe}. (Also notice that a coordinate-independent method to formulate the smoothly-crossing condition can be found in Ref. \cite{Hong:2022thd}.)

However, fixing the discontinuity at the thin shell in the Penrose diagram comes with a price. To see the problem, we calculate the metric in the modified KS coordinates $(V''_{+}, U''_{+})$.   From Eq.~(\ref{Second_conformal_BHtoWH}), one can easily check that the inverse transformation of the second conformal transformation exists and is given by  
\begin{equation}
(V'_{+}, U'_{+})=\left(V''_{+} \left|V''_{+}\right|^{\frac{1-k}{k}}, U''_{+} \left|U''_{+}\right|^{\frac{1-k}{k}}\right).
\end{equation}
Using Eq.~(\ref{Second_conformal_BHtoWH}) and the above relation, we can derive the differential relation 
\begin{equation}\label{Differential_relation_2nd_trans}
dV'_{+}=\frac{1}{k}  \frac{V'_{+}}{V''_{+}} dV''_{+}=\frac{1}{k} \left|V''_{+}\right|^{\frac{1-k}{k}} dV''_{+}.
\end{equation}
Next, the metric components in the original null KS coordinates $(V_{+}, U_{+})$ are given by 
\begin{equation}\label{Null_KS_metric_+}
d s_{+}^2=-\frac{16M_{+}^3}{r_{+}}e^{-r_{+}/2M_{+}}(d V_{+} d U_{+}+d U_{+} d V_{+})+r_{+}^2 d \Omega^2.
\end{equation}
Then using Eq.~(\ref{Differential_relation_2nd_trans}), we can show that after the two conformal transformations  
\begin{equation}
(V_{+}, U_{+})=(X_{+}V'_{+}, X_{+}U'_{+}) =\left(X_{+}V''_{+} \left|V''_{+}\right|^{\frac{1-k}{k}}, \; X_{+}U''_{+} \left|U''_{+}\right|^{\frac{1-k}{k}}\right),
\end{equation}
the metric components become
\begin{equation}\label{metric_after_2nd_trans}
d s_{+}^2=-\frac{16M_{+}^3}{r_{+}}e^{-r_{+}/2M_{+}}\frac{X_{+}^2}{k^2} \left|V''_{+}U''_{+}\right|^{\frac{1-k}{k}} (d V''_{+} d U''_{+}+d U''_{+} d V''_{+})+r_{+}^2 d \Omega^2, 
\end{equation}
which compared to Eq.~(\ref{Null_KS_metric_+}) contains an extra factor $\frac{X_{+}^2}{k^2}\left|V''_{+}U''_{+}\right|^{\frac{1-k}{k}}$. 
When $M_{+} < M_{-}$, which leads to $k \equiv \frac{M_{+}}{M_{-}}\sqrt{f_{+}(b)/f_{-}(b)} < 1$, the metric components $g_{U''_{+}V''_{+}}$ and $g_{V''_{+}U''_{+}}$ are zero at both the black hole horizon $(U''_{+}=0)$ and the white hole horizon  $(V''_{+}=0)$. Therefore, in the coordinate system $(V''_{+}, U''_{+}, \theta, \phi)$ if we calculate the norm of any four-vector with the components $T^{\mu}=(a, b, 0, 0)$ at the horizon, we will obtain zero as a result. However, in a well-behaved coordinate chart without singularity, only a null vector has a zero norm. Thus when $k<1$,  the metric (\ref{metric_after_2nd_trans}) has a new type of coordinate singularity causing a degeneracy of four-vectors with the form $T^{\mu}=(a, b, 0, 0)$ to null vectors. On the other hand, when $M_{+} > M_{-}$, \textit{i.e.} $k>1$, the metric components $g_{U''_{+}V''_{+}}$ and $g_{V''_{+}U''_{+}}$ diverge at the event horizon. This indicates another different type of coordinate singularity compared to the $k<1$ case.  

To see the difference, we look into the behavior of the trajectories of the radial timelike geodesics in Fig.~\ref{fig:second_trans_BHtoWH}, in which the upper half of each diagram is constructed by using the second conformal transformation (\ref{Second_conformal_BHtoWH}), but the lower half is not. While those trajectories behave as the usual timelike trajectories at the event horizons of the lower half of the diagrams, their behaviors around the event horizons of the upper half are drastically different. In the left figure corresponding to the divergent type of coordinate singularity ($k>1$), the trajectories of radial timelike geodesics become parallel to the event horizons at the very place. In contrast, in the right figure corresponding to the vanishing type of coordinate singularity ($k<1$), the trajectories of the same geodesics become transverse to the event horizons there. One can verify this result in the modified Penrose diagram, Fig.~\ref{fig:second_trans_BHtoWH}, by calculating the quantity $d\tilde{U}/d\tilde{V}$ with respect to a radial timelike geodesic. For convenience, we will call this quantity of a timelike geodesic the ``slope'' of its trajectory, since this name
is well-justified after we rotate the Penrose diagram $45\degree$ clockwise. Then, for instance, at the white hole event horizon of the upper half diagram, the slope of the trajectory of a radially outgoing timelike geodesic is given by
\begin{equation}\label{problem_at_horizon}
\left. \frac{d\tilde{U}}{d\tilde{V}} \right\vert_{r \rightarrow 2M_{+}}
=\left(1+\left(\frac{-U_{+}}{X_{+}}\right)^{2k}\right)\left(\frac{V_{+}}{U_{+}} \right)^{k-1} \frac{4eE_{+}^2}{U_{+}^2}, 
\end{equation}
where $E_{+}$ is the energy (per unit mass) corresponding to the geodesic. (The detailed derivation of this result is again given in the companion article focusing on the trajectories of the timelike radial geodesics \cite{Lin:2023sxe}.) In this form, we can clearly see that at the white hole event horizon $V_{+}=0$, the slope is either equal to zero if $k>1$, or divergent if $k<1$. That is, at the white hole event horizon, the trajectories become parallel to the horizon if $M_{+}>M_{-}$, but they become transverse to the horizon if $M_{+}<M_{-}$. Therefore, Eq.~(\ref{problem_at_horizon}) shows a new problem of the modified Penrose diagram for the black-to-white hole bouncing model with $M_{+} \neq M_{-}$, as the cost of fixing the discontinuity at the thin shell: a degeneracy of the four-velocities caused by a new type of coordinate singularity of the metric (\ref{metric_after_2nd_trans}) at the event horizons on the upper half of the diagram.

Lastly, in the special case $M_{+}=M_{-}$, the second conformal transformation (\ref{Second_conformal_BHtoWH}) is unnecessary or equivalent to an identity mapping, so there is no discontinuity in the resulting Penrose diagram using the simple cut-and-paste procedure mentioned in Sec. \ref{subSec: 1st_conformal_trans}. 
Since $M_{+}=M_{-}$ also leads to $dt_{+}=dt_{-}$ by Eq.~(\ref{t_relation_BWH}), which further leads to $E_{+}=E_{-}$ by Eq.~(\ref{Blue_shift_BWH}), we can also say that when there is no energy-shifting effect on the timelike geodesics crossing the thin shell, the second conformal transformation is not required. One then might have the inclination to deduce that for other spacetime models constructed via a static spacelike thin shell, once $dt_{+}=dt_{-}$ at the transition surface, \textit{i.e.} in the no-energy-shift cases, there is no discontinuity in the resulting Penrose diagram after the simple cutting-and-pasting procedure, so a second conformal transformation is not required. However, this deduction is simply incorrect. In the next section, we show that it is not such a case by considering the Schwarzschild-to-de Sitter transition through a static thin shell.     

\section{The Schwarzschild-to-de Sitter Transition}\label{Sec.BHtodS}

In this section, we study the similar issue in an asymmetric model constructed via thin shell approximation, the Schwarzschild-to-de Sitter transition proposed by Frolov,
Markov, and Mukhanov \cite{Frolov:1988vj}. Notice that in this scenario, the location of the transition surface $r_{\pm}=b$ is still the minimal radius of the Schwarzschild part, but it is the maximal radius of the de Sitter part. The stability issue of such a spacetime structure was studied by Balbinot and Poisson \cite{Balbinot:1990zz}, and they found it can be stable by tuning the equation of state of the shell. While in the original work, $b$ was assumed to be around the Planck length \cite{Frolov:1988vj}, here we do not make any assumption on the value of $b$ besides the condition that the thin shell is inside the black hole event horizon but outside the de Sitter event horizon of the reference observer at the south/north pole. In the following, we first review the static coordinate of de Sitter space and perform similar calculations as what we have done in Sec. \ref{Sec:BHtoWH}. We then show that a second transformation is also generally required; hence, a similar coordinate singularity is generated at the corresponding event horizon. (In Ref.~\cite{Frolov:1988vj}, the authors have suggested using the Kruskal type analytical continuation to describe the complete spacetime. Thus, this section can be viewed as a demonstration that the single Kruskal type of coordinate system fails to describe the complete spacetime. We thank an anonymous referee points this out to us.)

\subsection{The static coordinates of de Sitter space} 
To connect a black hole to de Sitter space, we work in the static coordinates of de Sitter space, in which the metric is given by 
\begin{equation}\label{dS_static_metric}
ds^2=-f_{dS}(r)dt^2+f_{dS}^{-1}(r)dr^2+r^2d\Omega^2, 
\end{equation}
with 
\begin{equation}
f_{dS}(r)=1-\frac{r^2}{\ell^2}, 
\end{equation}
where $\ell$ is the characteristic length scale of a de Sitter space, which is related to the Hubble constant $H$ and the cosmological constant $\Lambda$ by
\begin{equation}
\ell=\frac{1}{H}=\left(\frac{8 \pi }{3}\Lambda \right)^{-1/2}. 
\end{equation}
The static metric (\ref{dS_static_metric}) covers only a part of de Sitter space, the region with $r<\ell$, and there exists an event horizon at $r=\ell$ for the observer at rest at $r=0$. However, similar to the Schwarzschild solution, the metric can be extended to the region with $r>\ell$ provided that $r$ becomes the temporal coordinate and $t$ becomes a spatial coordinate. Moreover, analogous to Eq.~(\ref{radial_geodesic_Deq}), the four-velocity of an observer moving along with a radial timelike geodesic is given by
\begin{equation}\label{radial_geodesic_Deq_dS}
\mathcal{U}^{\alpha}=\left(\frac{d t}{d \tau}, \frac{d r}{d \tau}, \frac{d \theta}{d \tau}, \frac{d \phi}{d \tau}\right)=\left(\frac{\mathcal{E}}{f_{dS}}, -\epsilon \sqrt{\mathcal{E}^2-f_{dS}}, 0, 0\right), 
\end{equation}
where $\tau$  and $\mathcal{E}$ are the proper time and a conserved quantity associated with the geodesic, respectively. Also, we have $\epsilon=1$ for the radially ingoing geodesics and $\epsilon=-1$ for the radially outgoing geodesics.

A special choice of the conserved quantity $\mathcal{E}$ is $\mathcal{E}=1$. In such a case, the $(t, r)$-components in (\ref{radial_geodesic_Deq_dS}) reduces to 
\begin{equation}\label{dS_E=1}
\left(\frac{d t}{d \tau}, \frac{d r}{d \tau} \right)=\left(\frac{\ell^2}{\ell^2-r^2}, -\epsilon\frac{r}{\ell}\right), 
\end{equation}
which, for $r=0$, gives $dt/d\tau =1$ and $dr/d\tau=0$. That is, in the static coordinate the temporal coordinate $t$ is the proper time of the observer who is at rest at the origin.  Another special situation is when $\mathcal{E}=0$. Substituting $\mathcal{E}=0$ into (\ref{radial_geodesic_Deq_dS}) gives us 
\begin{equation}\label{dS_E=0}
\left(\frac{d t}{d \tau}, \frac{d r}{d \tau}, \frac{d \theta}{d \tau}, \frac{d \phi}{d \tau}\right)=\left(0, -\epsilon\sqrt{\frac{r^2}{\ell^2}-1}, 0, 0\right), 
\end{equation}
which has to satisfy $r \geq \ell$ to have real solutions. This means that in $r \geq \ell$ region, worldlines with constant $t$, $\theta$ and $\phi$ are geodesics. This special group of geodesics is the counterpart of the group of geodesics with $E=0$ comoving with respect to the Kantowski-Sachs spacetime inside the event horizon. We will utilize both of them when we perform the matching analysis analogous to the procedures used Sec. \ref{subSec: 1st_conformal_trans}.   

\begin{figure}[h!]
	\centering
	\includegraphics[scale=0.3]{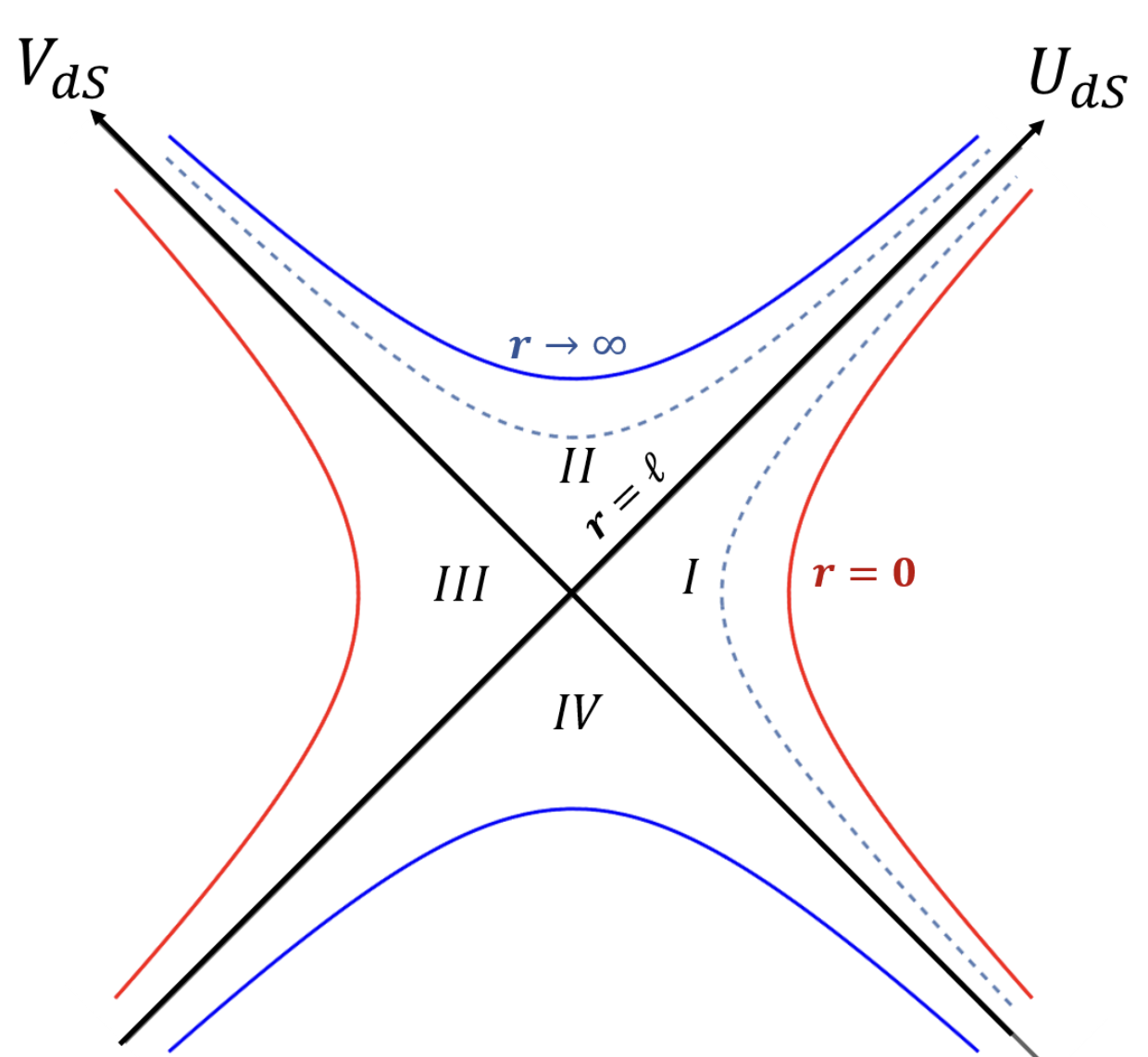}
	\caption{ The Kruskal diagram of a de Sitter space. Notice the difference of the convention of $(U_{dS},V_{dS})$ here and $(V, U)$ in Fig~\ref{fig:Simple_KS}.}  
	\label{fig:deSitter_Kruskal}
\end{figure}

Next, the coordinate singularity at the de Sitter horizon $r=\ell$ can be removed by introducing the retarded and advanced coordinates: $u_{dS}=t-r^*_{dS}$ and $v_{dS}=t+r^*_{dS}$, where $r^*_{dS}$ is the  
corresponding tortoise coordinate 
\begin{equation}\label{tortoise_dS}
r^*_{dS}= \frac{\ell}{2} \log \left|\frac{\ell+r}{\ell-r}\right|. 
\end{equation}
Then, one can introduce the KS-like null coordinates for de Sitter space \cite{Spradlin:2001pw, Zee:2013dea}
\begin{equation}\label{KS_deSitter}
 \left(U_{dS}, V_{dS} \right) \equiv \left(\pm  e^{u_{dS}/\ell} , \pm  e^{-v_{dS}/\ell} \right),  
\end{equation}
where the plus/minus signs are determined by the quadrants considered in the corresponding Kruskal diagram of de Sitter space, see Fig.~\ref{fig:deSitter_Kruskal}, and $(U_{dS}, V_{dS})$ satisfy the following relations
\begin{equation}\label{UV_dS}
U_{dS}V_{dS}=\frac{r-\ell}{r+\ell},
\end{equation}
and
\begin{equation}\label{U/V_dS}
\frac{U_{dS}}{V_{dS}}=\pm e^{2t/\ell}.     
\end{equation}
In the Kruskal diagram, the trajectory  $U_{dS}/V_{dS}=\pm e^{2t_a/\ell}$ with some constant $t_a$ belongs to the family of geodesics with $\mathcal{E}=0$ in $r \geq \ell$ region. By using the KS-like null coordinates from Eq.~(\ref{KS_deSitter}), 
the KS form of the de Sitter metric is given as
\begin{equation}\label{metric_de_Sitter_Kruskal}
\begin{split}
ds^2=&\ell^2 \left[\frac{-4}{(1-U_{dS}V_{dS})^2}dU_{dS}dV_{dS} + \frac{(1+U_{dS}V_{dS})^2}{(1-U_{dS}V_{dS})^2}d\Omega^2 \right], \\
=& -(r+\ell)^2 dU_{dS}dV_{dS} + r^2 d\Omega^2.
\end{split}
\end{equation}

\subsection{The Penrose diagram of the Schwarzschild-to-de Sitter transition}\label{Subsec:1st_trans_BHtodS}

Similar to Eq.~(\ref{induced_metric_relation_BWH}), in the Schwarzschild-to-de Sitter transition model via a static thin shell, the first junction condition  gives the following relation
\begin{equation}\label{induced_metric_relation_BH_dS}
-fd t_{-}^2+r_{-}^2d\Omega^2=-f_{dS}d t_{+}^2+r_{+}^2d\Omega^2, 
\end{equation}
where $f(r_{-})=1-2M/r_{-}$ and $f_{dS}(r_{+})=1-r_{+}^2/\ell^2$. 
Thus we have $r_{-}=r_{+}=b$, and 
\begin{equation}\label{1JC_Sch_to_dS}
\sqrt{\frac{2M}{b}-1}\;dt_{-}=\sqrt{\frac{b^2}{\ell^2}-1}\;dt_{+}, 
\end{equation}
which indicates that no-energy-shift, $dt_{dS}/dt_{BH} = 1$, requires 
\begin{equation}
b=\left(2M\ell^2\right)^{1/3}.
\end{equation}
\begin{figure}[h!]
	\centering
	\includegraphics[scale=0.5]{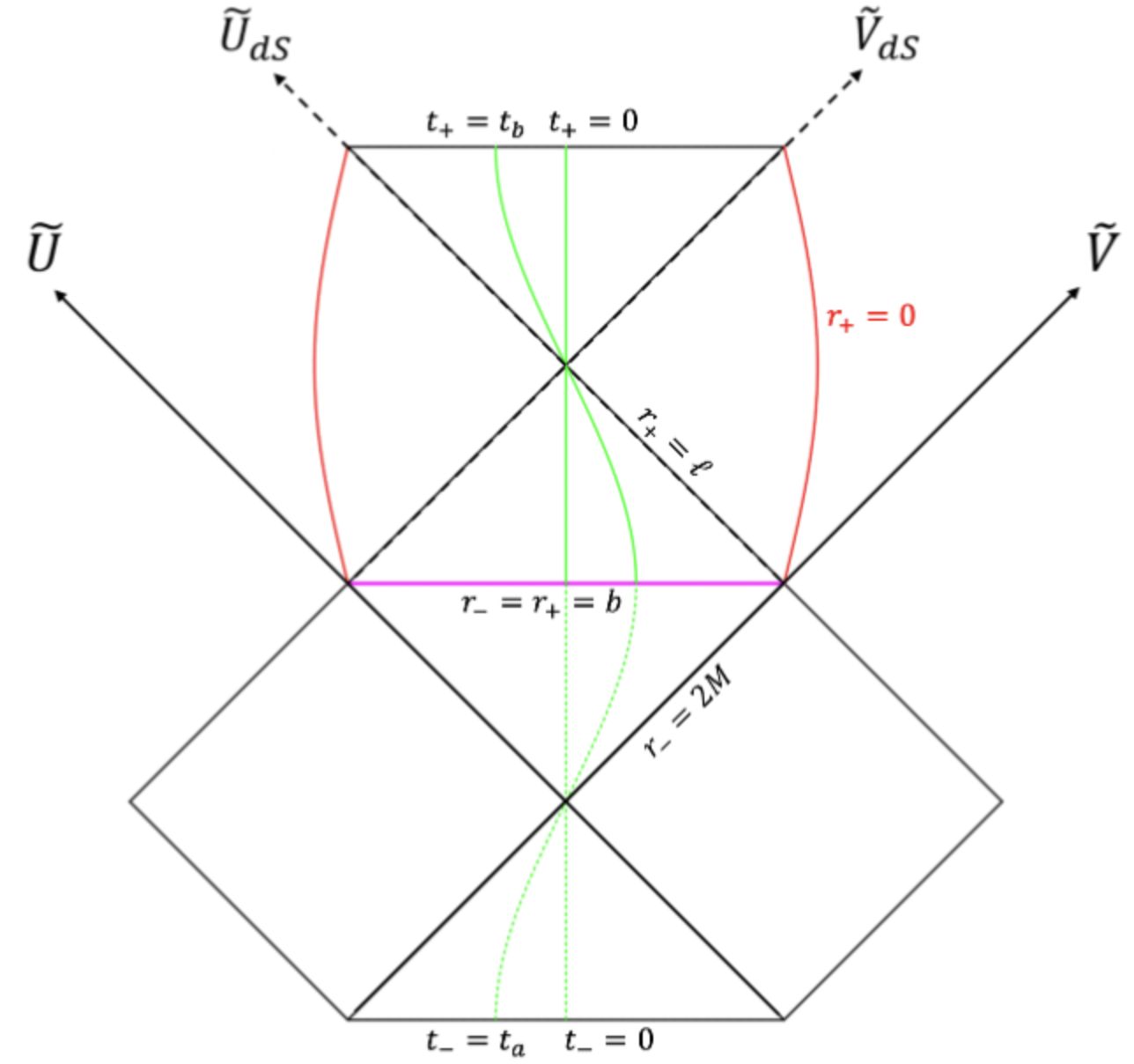}
	\caption{ The cut-and-pasting result after the rescaling transformation (\ref{BH_dS_rescaling_trans}). Similar discontinuity happens at the thin shell as that in the black-to-white hole transition case, Fig.~\ref{fig:first_trans_result}. The seemingly connected green trajectories corresponding to special geodesics with constant $t_{\pm}$ are, in fact, different geodesics except for the reference one with $t_{\pm}=0$.  }  
	\label{fig:BH_dS_1st_matching}
\end{figure}

To construct the corresponding Penrose diagram without the unwanted regions: $r_{-}<b$ and $r_{+}>b$, the procedure for the black hole part is given in  Sec. \ref{subSec: 1st_conformal_trans}, while the procedure for the de Sitter phase is quite similar as follows. To remove the unwanted region $r_{+}>b$ in de Sitter space, we first perform the rescaling transformation 
\begin{equation}\label{BH_dS_rescaling_trans}
(U'_{dS}, V'_{dS})=\left(\frac{U_{dS}}{X_{dS}}, \frac{V_{dS}}{X_{dS}}\right), 
\end{equation}
where 
\begin{equation}
X_{dS}^2 \equiv \frac{b-\ell}{b+\ell},
\end{equation} 
together with the inverse tangent transformation  
\begin{equation}
(\tilde{V}_{dS}, \tilde{U}_{dS})= \left(\tan^{-1}V'_{dS}, \tan^{-1}U'_{dS}\right). 
\end{equation}
With a similar cutting-and-pasting procedure and the following identification for the de Sitter phase 
\begin{equation}\label{BH_dS_connection_rule}
(\tilde{V}, \tilde{U}) =
     \left(\tilde{V}_{dS}+\frac{\pi}{2}, \tilde{U}_{dS}+\frac{\pi}{2}\right),    
\end{equation} 
we obtain the Penrose diagram given by the simple cut-and-paste procedure, Fig.~\ref{fig:BH_dS_1st_matching}. (Similar to Fig.~\ref{fig:first_trans_result}, here is the intermediate result of the first transformation, which requires further transformation.)

Again, there is a consistent issue we have to check by matching the trajectories of geodesics with constant $t$ in the black hole part, Eq.~(\ref{KS_Ober_in_minus}), with their de Sitter space counterpart at the thin shell. In de Sitter space, a geodesic with constant ``position'' $t=t_b$ described by Eq.~(\ref{U/V_dS}) with four-velocity (\ref{dS_E=0}) intersects the thin shell at point  
\begin{equation}
\begin{split}\label{dS_geodesics_thin_shell}
(\tilde{V}, \tilde{U} )&= \left(\tilde{V}_{dS}+\frac{\pi}{2}, \tilde{U}_{dS}+\frac{\pi}{2}\right)  \\
&=\left(\frac{\pi}{2}-\tan^{-1} e^{-t_b/\ell}, \frac{\pi}{2}-\tan^{-1} e^{t_b/\ell}\right) \\
&= \left(\tan^{-1} e^{t_b/\ell}, \tan^{-1} e^{-t_b/\ell}\right).
\end{split}
\end{equation}
Then at the thin shell, matching  Eq.~(\ref{dS_geodesics_thin_shell}) from the de Sitter phase to Eq.~(\ref{KS_crossing_from_minus}) from the Schwarzschild phase leads to the the identification
\begin{equation}\label{Sch_dS_relation_by_matching}
\frac{t_b}{\ell}=\frac{t_a}{4M},    
\end{equation}
which is generally different from the relation derived from the first junction condition (\ref{1JC_Sch_to_dS}): 
\begin{equation}\label{Sch_dS_relation_by_1JC}
\sqrt{\frac{2GM}{b}-1}\;t_a=\sqrt{\frac{b^2}{\ell^2}-1}\;t_b.  
\end{equation}
Therefore, there is also an implicit discontinuity at the thin shell in Fig.~\ref{fig:BH_dS_1st_matching}. However, the situation here is different from that of the generalized black-to-white hole bouncing model. Previously, when the masses of the black hole and white hole are the same, which corresponds to the no-energy-shift case, both Eqs.~(\ref{t_relation_BWH}) and (\ref{continuity_condition}) become $t_a=t_b$ so there is no discontinuity at the shell. Here, in the no-energy-shift case, \textit{i.e.} when $b=(2M\ell^2)^{1/3}$,  while Eq.~(\ref{Sch_dS_relation_by_1JC}) reduces to $t_a=t_b$, Eq.~(\ref{Sch_dS_relation_by_matching}) is still the same. Then, in this model, a second conformal transformation is still required to fix this mismatch, even if there is no energy shift. 

Nevertheless, we still can ask if there is always a natural radius $b=b(\ell, M)$ such that the conditions (\ref{Sch_dS_relation_by_matching}) and (\ref{Sch_dS_relation_by_1JC}) are the same, so the second conformal transformation is not required. Such a radius is given by the solution of the following equation
\begin{equation}
b^3+[(4M)^2-\ell^2]b-32M^3=0,
\end{equation}
which is derived by rewriting Eqs.~(\ref{Sch_dS_relation_by_matching}) and (\ref{Sch_dS_relation_by_1JC}) into the form with $t_a/t_b$ on one side of the equations. Substituting $\beta \equiv b/2M $, we can rewrite the equation into 
\begin{equation}
\beta^3+\left[4-\left(\frac{\ell}{2M}\right)^2\right]\beta-4= 0.
\end{equation}
Within the range $0<\ell/2M <1$, this equation always have one real solution given by
\begin{equation}\label{special_b_value_BHtodS}
\beta=\frac{-4+(\ell/2M)^2}{\left(54+\sqrt{54^2+(12-3(\ell/2M)^2)^3}\right)^{1/3}}+\frac{1}{3}\left(54+\sqrt{54^2+(12-3(\ell/2M)^2)^3}\right)^{1/3}, 
\end{equation}
which increases monotonically from $\beta \sim 0.8477$ when $\ell/2M \rightarrow 0$, to $\beta=1$ when $\ell/2M \rightarrow 1$ and the natural condition of the model $\beta>\ell/2M$ is satisfied always. Also, in any case, this natural radius is comparable to the size of the radius of the black hole event horizon. 

\begin{figure}[h!]
	\centering
	\includegraphics[scale=0.8]{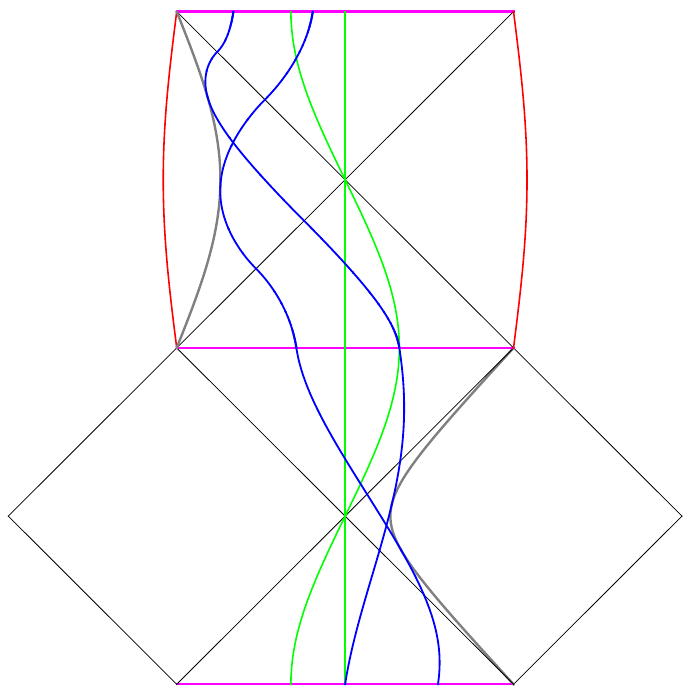}
	\caption{The Penrose diagram of the a BH-to-de Sitter transition with the shell located at radius $b=2M \times 7/10$, and the de Sitter scale $\ell=2M \times 1/6 $.  In this resulting Penrose diagram, though there is no discontinuity at the thin shell, timelike radial geodesics (blue) become transverse to the de Sitter event horizon (squeezing type, see  Sec. \ref{Subsec: geometric interpretation}), due to the coordinate singularity generated by the second transformation (\ref{Second_conformal_BH_dS}). }  
	\label{fig:Sch_to_deSitter}
\end{figure}

To fix the discontinuity at the shell in the corresponding Penrose diagram, one can perform a second transformation on either the Schwarzschild phase or the de Sitter phase. Since we have worked out its results and have discussed its issues on the Schwarzschild phase in Sec. \ref{subSec:second_trans_BHtoWH}, here, as a demonstration, we choose to work out its result on the de Sitter phase.   

To enforce the relation (\ref{Sch_dS_relation_by_1JC}), we perform a second conformal transformation on $(V'_{dS}, U'_{dS})$ as
\begin{equation}\label{Second_conformal_BH_dS}
\left(V''_{dS}, U''_{dS}\right)=\left(V'_{dS} \left| V'_{dS} \right| ^{k_{dS}-1}, U'_{+} \left| U'_{+} \right| ^{k_{dS}-1} \right),
\end{equation}
where 
\begin{equation}
    k_{dS} \equiv \frac{\ell}{4M}\sqrt{\frac{f_{dS}(b)}{f_{BH}(b)}},
\end{equation}
such that 
\begin{equation}\label{second_conformal_dS_t}
\frac{U''_{dS}}{V''_{dS}}=\pm e^{\frac{t_{dS}}{2M}\sqrt{f_{dS}(b)/f_{BH}(b)}}. 
\end{equation}
With this extra transformation, Eq.~(\ref{Second_conformal_BH_dS}), a modified Penrose diagram without discontinuity around the thin shell can be constructed as shown in Fig.~\ref{fig:Sch_to_deSitter}.  Again, to fix the discontinuity, what we sacrificed is the good property of the original coordinate system at the de Sitter event horizon. Using the second form of Eq.~(\ref{metric_de_Sitter_Kruskal}) and Eq.~(\ref{Second_conformal_BH_dS}), one can derive the metric in coordinates $(U''_{dS}, V''_{dS}, \theta, \phi )$ as 
\begin{equation}\label{dS_metric_after_2nd_trans}
d s_{dS}^2=-(r_{+}+\ell)^2\frac{X_{dS}^2}{k_{dS}^2} \left|V''_{dS}U''_{dS}\right|^{\frac{1-k_{dS}}{k_{dS}}} (d V''_{dS} d U''_{dS}+d U''_{dS} d V''_{dS})+r_{+}^2 d \Omega^2.  
\end{equation}
Similar to the discussion under Eq.~(\ref{metric_after_2nd_trans}), if $k_{dS} \neq 1$, the factor $\left|V''_{dS}U''_{dS}\right|^{\frac{1-k_{dS}}{k_{dS}}}$ causes a coordinate singularity at the de Sitter event horizon ($V''_{dS}=0$ or $U''_{dS}=0$), which leads to the distortion of the trajectories of the timelike radial geodesics at the de Sitter horizon as shown in Fig.~\ref{fig:Sch_to_deSitter}. 

Thus, we see that in this asymmetric example, mismatching is also generated by the simple cut-and-paste procedure. And to solve it, again, we must use a second transformation on one of the two spacetime manifolds connected via the thin shell. This action, nevertheless, brings back a coordinate singularity at the event horizon. So why does the second transformation unavoidably reintroduce a special type of coordinate singularity back to the event horizon? Also, what is the general criterion for the resulting Penrose diagram of the simple cut-and-paste procedure to be automatically free from discontinuity at the thin shell? The answers are, in fact, rooted in the geometric features of the Penrose diagrams, which we will discuss in the next section.

\section{Discussions}\label{Sec:discussions}

\subsection{The geometric interpretation of the coordinate singularity}\label{Subsec: geometric interpretation}

\begin{figure}[h!]
	\centering
	\includegraphics[scale=0.7]{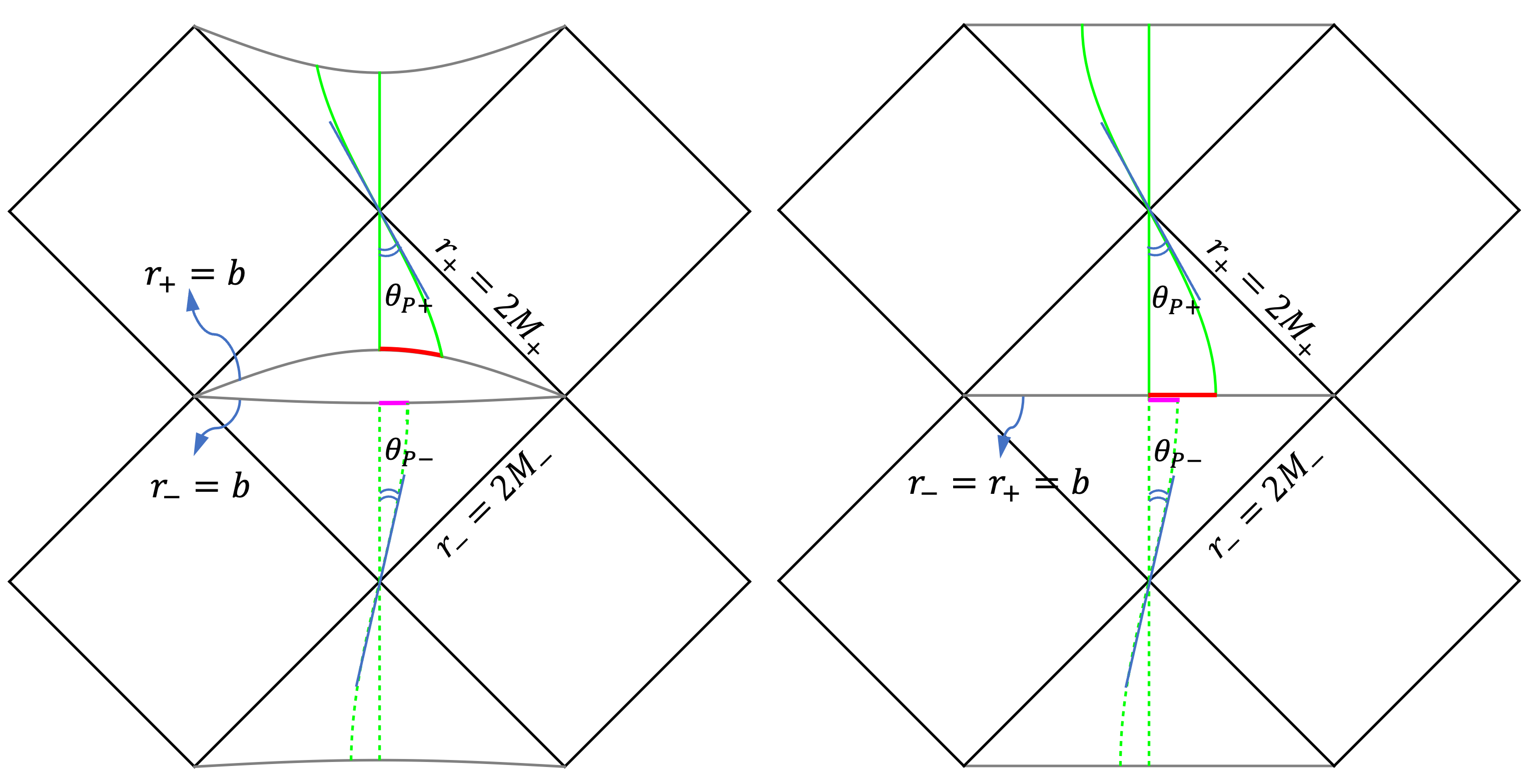}
	\caption{ The Penrose diagrams of a generalized black-to-white hole bounce before (left) and after (right) the rescaling transformation (\ref{ModKS}). The lines in red and magenta are the visual length of half of the rod (from its center of mass to one of its ends). They are generally not the same on both of the diagrams due to the fact that the Penrose diagrams are normalized such that the length of an event horizon we see on a diagram is of unity. }  
	\label{fig:the_meaning_of_singularity}
\end{figure}

Before we offer a geometrical explanation of the coordinate singularities in Eqs.~(\ref{metric_after_2nd_trans}) and (\ref{dS_metric_after_2nd_trans}), we would like to clarify two possible questions regarding the second transformation enforcing the first junction condition.
Also, in this subsection, we use the generalized black-to-white hole bouncing model as an example for demonstration, but the argument is general to spherical symmetric spacetime models constructed by a static thin shell.

Firstly, one might wonder if a rescaling transformation $t_{+}=kt'_{+}$ with $k \equiv \frac{M_{+}}{M_{-}}\sqrt{f_{+}(b)/f_{-}(b)}$ can achieve the same effect of the transformation (\ref{Second_conformal_BHtoWH}). The answer to this question is definitely not. The effect of $t_{+}=kt'_{+}$ is totally different than the effect of Eq.~(\ref{Second_conformal_BHtoWH}).  Using $t_{+}=kt'_{+}$, what one will have is  $U'_{+}/V'_{+}=\pm \exp{[-\sqrt{f_{+}(b)/f_{-}(b)} \;t'_{+}/2M_{-}]}$ instead of Eq.~(\ref{second_conformal_moving_t}). At the same time the rescaling $t_{+}=kt'_{+}$ also changes the relation from the first junction condition accordingly, so it can never solve the problem of discontinuity. In fact, a rescaling $t_{+}=kt'_{+}$ is just a change of unit, like putting the symbol of the speed of light $c$ back to our expression!  

The second question one might have is: Is it possible to have a better second transformation such that it fixes the discontinuity at the thin shell, but its effect decreases to zero at the event horizon? The answer to this question is also negative based on our purpose here. If such a transformation exists, it must be $r$-dependent. However, in both models we have considered, $r$ is a function of the form: $r=r(UV)$, where $(U, V)$ are the corresponding KS coordinates. To preserve the trajectory of null geodesics at $45\degree$, only transformations of the type given in Eq.(\ref{preserving_45_trans}) are allowed, so such a transformation cannot be used to construct a Penrose diagram. Simply put, since generally $V'(r(UV)) \neq V'(V)$, there is no better second transformation that preserves the sacred property of a Penrose diagram and solves the issue of discontinuity at the same time. However, a global chart for the spacetime is possible by using this type of $r$-dependent transformation, and we will discuss this later.

Now, to construct the geometrical explanation, we first notice that the Penrose diagram is normalized such that the length of an event horizon we see on a diagram is of unity. Therefore, the physical scale related to the event horizon is hidden from the diagram. A similar analogy to this point (and only regarding this point) is like representing all circles by unit circles. Then, if we know the original length of the radius $R$, a given length $L$ on the original circle can be represented by the angle corresponding to the arc $\theta=L/R$. Thus, the larger the original radius is, the smaller the angle and arc we see on the corresponding unit circle.

In a Penrose diagram, the relation between the angle and a physical length at a given radius $r=b$ is more complicated and depends on the location. Nevertheless, as mentioned in Sec. \ref{subSec:BHtoWH_model}, a comparison between the upper and lower part of the Penrose diagram, as shown in the left figure in Fig.~\ref{fig:the_meaning_of_singularity}, can be made by 
the following procedure. We consider a uniform rigid rod with its center of mass moving along with the radial geodesic specified by $t_{\pm}=0$, and its body aligned along the $t$-direction.  When the rod crosses the thin shell at $r_{\pm}=b$, the positions where the worldlines of its center of mass and one of its ends intersect the thin shell define two spacetime events: event A and event B respectively. 
Though the end of the rod does not move along with any geodesic, we can set up another test particle whose worldline intersects the spacetime event B, while it moves along with the special radial geodesic specified by some $t_{\pm}=const$. Then we can define the angle where the worldline of the rod's center of mass and the worldline of the test particle meet in the Penrose diagram as $\theta_{P \pm}$. Then it is not difficult to show that for a rod with physical length $2 L_{phy}$ (so from the center of mass to one of its ends is $L_{phy}$), the corresponding angles in the unmodified Penrose diagrams $\theta_{P \pm}$ is given by 
\begin{equation}\label{angle_phys_L_relation}
\theta_{P\pm}=\left|\frac{\pi}{4}-\tan^{-1}\left[e^{-L_{phy}/(2M_{\pm}\sqrt{-f_{\pm}(b)})}\right]\right|.
\end{equation}
Although the relation (\ref{angle_phys_L_relation}) is more complicated than the unit circle analogy, $\theta=L/R$, in the black-to-white-hole bouncing model, we still have the result that the larger the mass parameter, the smaller the angle. (On the other hand, regarding this point in the Schwarzschild-to-de Sitter case, the transition radius $r=b$ enters this physical-length-to-angle relation non-trivially.
Nevertheless, the argument made in this section can still be directly applied. ) Also, the rod's visual length, which we see directly on the diagram, is of a positive monotonic relationship with the angle $ \theta_{P \pm}$. That is, the larger the angle, the longer the visual length. 

Now, while the rescaling transformation (\ref{ModKS}) brings the shell into the same curves in the two coordinate charts, it also keeps the angle $\theta_{P \pm}$ unchanged. Therefore, we end up with a figure with one shell with two different visual lengths for a single rod on it, as shown in the right figure in Fig.~\ref{fig:the_meaning_of_singularity}.  
To eliminate this issue, we need to use a second transformation, $\theta_{P \pm} \rightarrow \theta'_{P \pm}$, either to expand the angle on the side with a larger mass or to shrink the angle on the side with a smaller mass. However, the second transformation also keeps the null geodesics at $45\degree$ and the position of the event horizons unchanged, so the lines for event horizons still cross each other transversely in the resulting diagram. That is, the maximum value of $\theta'_{P \pm}$ is still $45\degree$. Keeping the maximal $\theta'_{P \pm}$ unchanged but increasing or decreasing the angles corresponding to any finite physical length causes the singularity at the maximal $\theta'_{P \pm}$, \textit{i.e.} at the event horizons, in the modified Penrose diagram. This geometric explanation is manifested in the behavior of the general radial geodesics in Fig.~\ref{fig:second_trans_BHtoWH}. In the stretching type of singularity, the trajectory of a radial geodesic at the event horizon is being stretched to become parallel to the event horizon. In contrast, in the squeezing type of singularity, the trajectory of a radial geodesic is squeezed to become transverse to the event horizon.  

Although we have argued that there is no better second conformal transformation that fixes the discontinuity issue without causing singularity at the related event horizons, we cannot rule out the possibility that one can remove the above-mentioned singularity by applying a third conformal transformation on the resulting Penrose diagram, \textit{e.g.} Fig.~\ref{fig:second_trans_BHtoWH}. In Sec.~\ref{Subsec_3rdTrans}, we discuss this third transformation and show that if we restrict to the simple $r$-dependent conformal coordinates, in which the metric components can be expressed as functions of $r$, no third conformal transformation of this type can make the metric regular at all event horizons. Nevertheless, more complicated conformal transformations can remove coordinate singularity at all event horizons while keeping the continuity at the shell mended by the second transformation. We discuss the construction of them in Sec.~\ref{Subsec_3rdTrans}.

Restricted to only the simple $r$-dependent conformal charts, this type of singularity can only be avoided when the angles $\theta_{P \pm}$ corresponding to a given physical length on the shell are the same. In the scenario where the spacetimes on the two sides of the shell are given by the same type of solutions, this coordinate singularity does not exist when all of the parameters specifying the solutions are the same. 
For instance, in the generalized black-to-white hole bouncing model or in an analogous de Sitter-to-de Sitter transition model, $\theta_{P+}=\theta_{P-}$ when $M_{+}=M_{-}$ or $\ell_{+}=\ell_{-}$ respectively. (In more complicated cases carrying more than one physical parameter, there might be more than one possible choice of parameters making $\theta_{P+}=\theta_{P-}$. However, setting all physical parameters on two sides of the shell to be the same should always make this type of singularity disappear. )
In contrast, when the two sides of spacetime are described by different types of solutions, for instance, the Schwarzschild-to-de Sitter transition studied in this work, for a given set of mass and de Sitter scale, the condition
$\theta_{P+}=\theta_{P-}$ depends on a special choice of transition surface $r=b$ given by Eq.~(\ref{special_b_value_BHtodS}). 

Lastly, if our goal is to construct the Penrose diagram properly, having conformal coordinate patches covering the entire shell is sufficient. To better understand this point, we compare the situation here to the physical Reissner-Nordstr{\"o}m solution, in which two horizons with different sizes exist.

\subsection{A comparison to the Reissner-Nordstr{\"o}m solution}\label{Subsec:RN}

\begin{figure}[h!]
	\centering
	\includegraphics[scale=0.5]{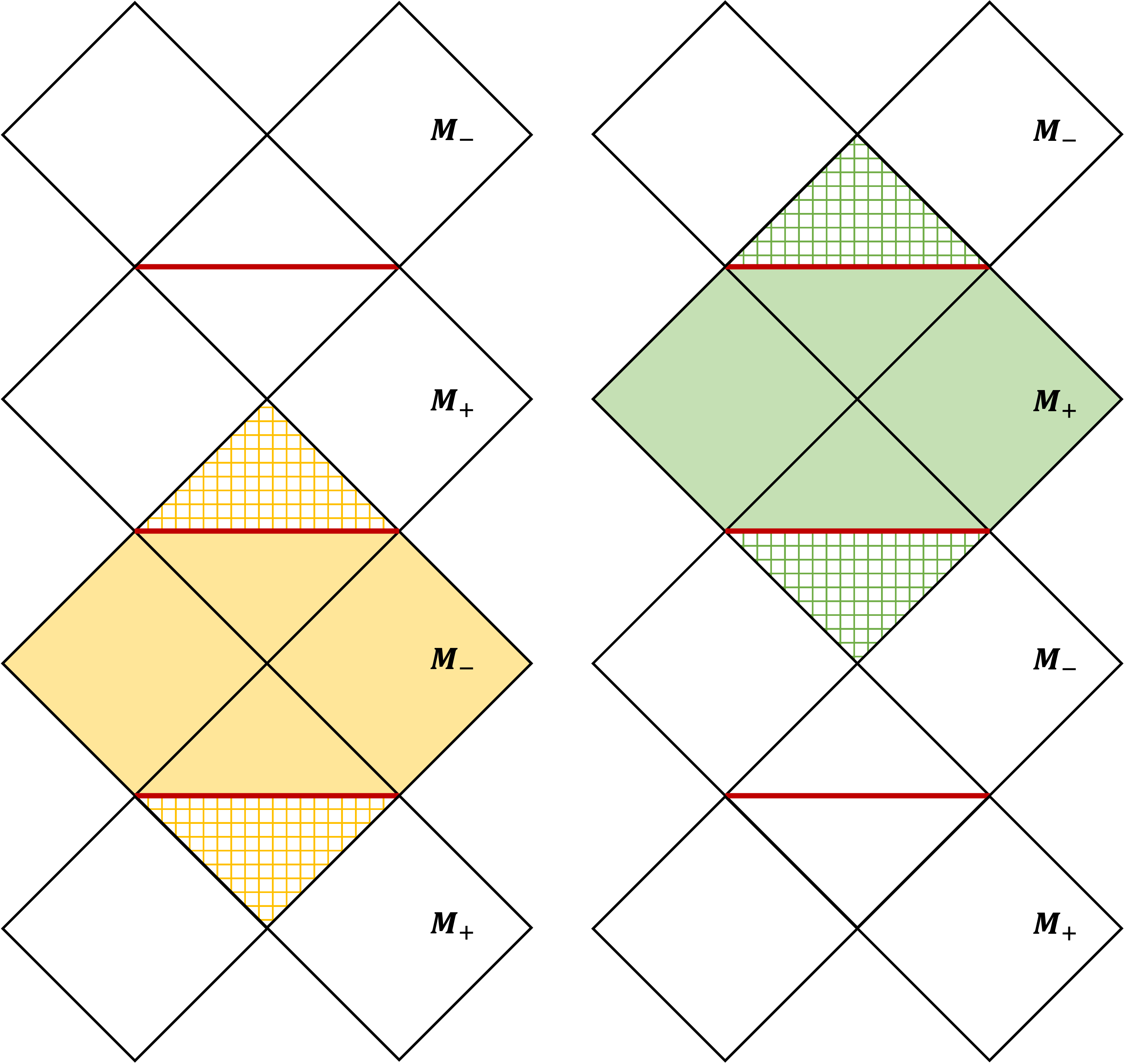}
	\caption{In the generalized black-to-white hole bouncing scenario, a final Penrose diagram should be constructed as the method used for the Penrose diagram of the RN solution. The parts filled with solid colors (yellow and green) are constructed through only the first transformation (\ref{ModKS}), while the parts filled with grid patterns (yellow and green) are constructed with the additional second transformation (\ref{Second_conformal_BHtoWH}). Then the patch combined with solid and grid parts, as shown in the left and right figures respectively, forms a well-defined coordinate patch covering the entire thin shell. A final Penrose diagram is then constructed by combining the yellow and green patches with the shell in the overlapping region of the two charts. Notice that without the second transformation, we only have the solid-colored parts, and the combining process is, in principle, unjustified since there is no overlapping between the coordinate charts. }  
	\label{fig:tiling}
\end{figure}

When $k<1$ and $k_{dS}<1$, the coordinate singularities found in both Eqs.~(\ref{metric_after_2nd_trans}) and (\ref{dS_metric_after_2nd_trans}) are similar to the situation in the \textbf{Reissner-Nordstr{\"o}m} (RN) solution. In the physical RN solution, \textit{i.e.} solution with $M^2-e^2>0$, there exist an outer horizon $r_{1}=M+\sqrt{M^2-e^2}$ and an inner horizon $r_{2}=M-\sqrt{M^2-e^2}$.  Then in terms of the null KS-like coordinates covering the outer horizon $(U_{1}, V_{1})$, the metric is given by \cite{Griffiths:2009dfa}
\begin{equation}\label{RN_metric_in_outer}
d s_{1}^2= \frac{-r_{1}r_{2}}{2r^2}\left(\frac{r-r_{2}}{r_{2}}\right)^{1+\frac{r^2_{2}}{r^2_{1}}} \exp\left[\frac{-(r_{1}-r_{2})r}{r^2_{1}}\right] (d V_{1} d U_{1}+d U_{1} d V_{1})+r^2 d \Omega^2.
\end{equation}
Thus the factor $(\frac{r-r_{2}}{r_{2}})^{1+\frac{r^2_{2}}{r^2_{1}}}$ makes the same type of coordinate singularity at the inner event horizon $r=r_{2}$ as the factor $\left|V''_{+}U''_{+}\right|^{\frac{1-k}{k}}$ and $\left|V''_{dS}U''_{dS}\right|^{\frac{1-k_{dS}}{k_{dS}}}$
in Eqs.~(\ref{metric_after_2nd_trans}) and (\ref{dS_metric_after_2nd_trans}) respectively. 
Similarly, in terms of the null KS-like coordinates covering the inner horizon $(U_{2}, V_{2})$, the RN metric becomes singular at the outer horizon $r=r_{1}$ due to a factor $(\frac{r-r_{1}}{r_{1}})^{1+\frac{r^2_{1}}{r^2_{2}}}$. 

Without a global conformal coordinate chart, nevertheless, the Penrose diagram of a physical RN solution can be constructed by compactifying the two coordinate systems and piling them alternatingly with overlapping in the $r_{2}<r<r_{1}$. 
Thus, for the models considered in this work, even without a global conformal coordinate chart, their final Penrose diagrams can still be constructed in a similar way. We emphasize that the Penrose diagram has to be constructed by coordinate charts with overlapping, which is a property lacking before we use the second transformation to fix the discontinuity at the shell. An example is given in Fig.~\ref{fig:tiling}. Nevertheless, in Sec.~\ref{Subsec_3rdTrans}, we construct a formalism for the third transformation built upon the result of the second transformation. Using this formalism, we then offer a conformal coordinate chart in which the metric components are regular at all event horizons. 

\subsection{The third conformal transformation preserving the continuity of coordinates at the shell }\label{Subsec_3rdTrans}

\begin{figure}[h!]
	\centering	
         \includegraphics[scale=0.16]{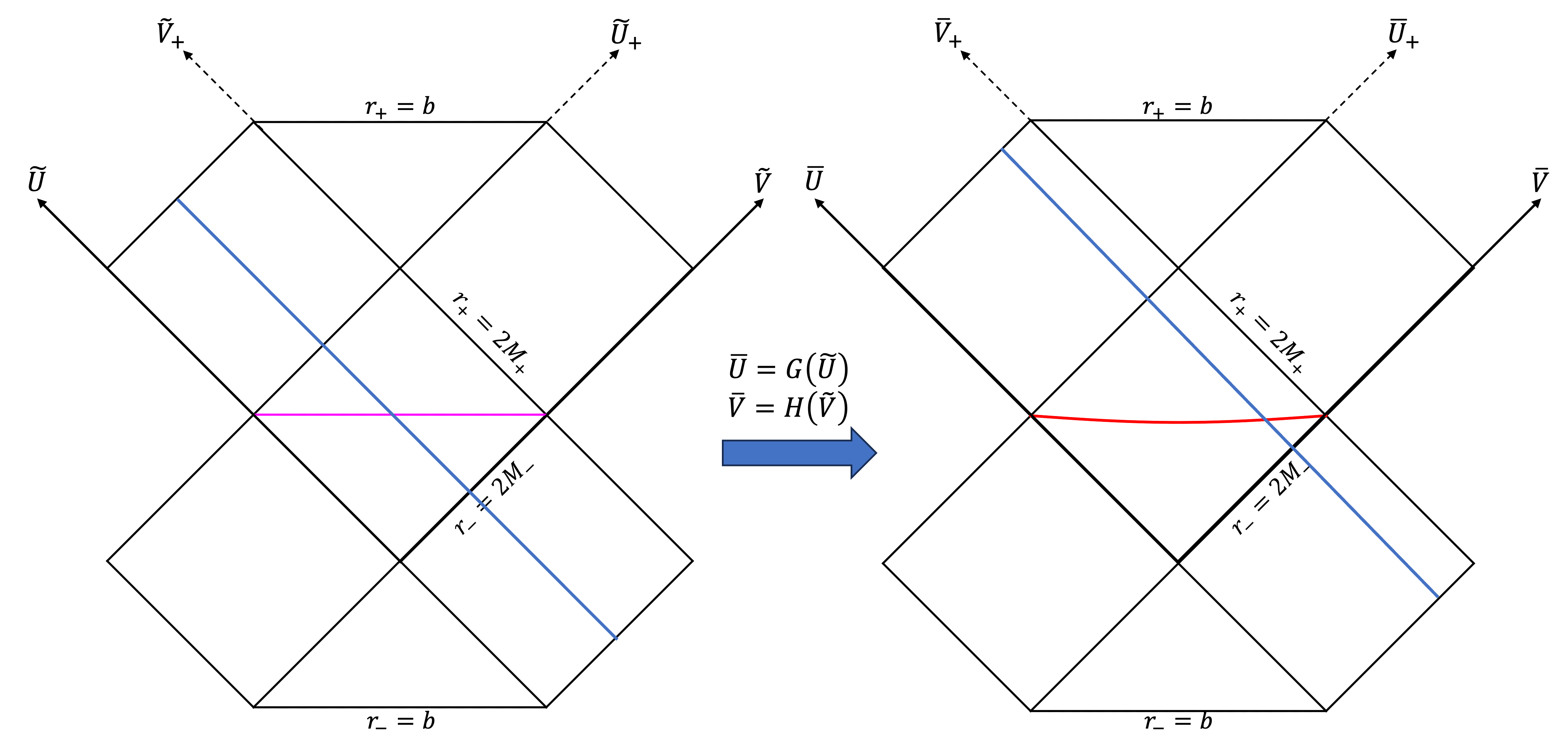}
	\caption{The conformal transformation preserving the continuity at the thin shell can be formulated by the transformations between the compactified diagrams. After the transformation Eq.~(\ref{CT_of_the_final_Penrose_D}), a null geodesic (blue), specified by $\tilde{V}=\tilde{V}_a$, crossing the thin shell (magenta) on the left diagram must also be a single straight line, specified by $\bar{V}=H(\tilde{V}_a)$, crossing the thin shell (red) on the right diagram. The 
 transformation Eq.~(\ref{CT_of_the_final_Penrose_D}) can be rewritten into the forms for the non-compactified coordinates: Eqs.~(\ref{CT_Penrose_lower}) and (\ref{CT_Penrose_upper}), for which the condition of the continuity at the thin shell leads to the relation Eq.~(\ref{CT_relation}).  }  	\label{fig:CT_of_Penrose_D}
\end{figure}

In Sec.~\ref{Subsec: geometric interpretation}, we have established the fact that no conformal transformation acting only on the $+$ region can fix the discontinuity while keeping the metric regular at the event horizon. Nevertheless, we have not exhausted all the possible conformal transformations on the models considered in this work. In this subsection, we consider the conformal coordinate transformation acting on both the upper and lower regions (the regions covered by the coordinates with subscript $+$ and $-$, respectively) after fixing the discontinuity issue at the thin shell.

After constructing the general formalism for such a conformal transformation, 
we first show that there is no \textbf{simple $r$-dependent} global conformal chart for the spacetime model constructed through a static spacelike thin shell. Then, we provide a global conformal coordinate system in which a regular metric can be defined everywhere by taking the limit value of the metric function at the event horizon. Here, what we mean by a simple $r$-dependent conformal chart is that in such a coordinate system, the metric component can be expressed only in terms of $r$, where $r$ is a function of the conformal coordinates used in the coordinate chart. This is the case for both the Kruskal coordinates of the Schwarzschild solution Eq.~(\ref{Kruskal_in_U_V}) and of the RN solution Eq.~(\ref{RN_metric_in_outer}). Notice that the metrics of the two examples discussed in Sec.~\ref{Sec:BHtoWH} and Sec.~\ref{Sec.BHtodS} also belong to this category.  For instance,  the metric components of the upper region of the generalized black-to-white hole bounce given by Eq.~(\ref{metric_after_2nd_trans}) can be rewritten into 
\begin{equation}\label{metric_after_2nd_trans_in_r}
d s_{+}^2=-\frac{16M_{+}^3}{r_{+}}e^{-r_{+}/2M_{+}}\frac{X_{+}^{2k}}{k^2} \left|\left(1-\frac{r_{+}}{2M_{+}}\right)e^{r_{+}/2M_{+}}\right|^{1-k} (d V''_{+} d U''_{+}+d U''_{+} d V''_{+})+r_{+}^2 d \Omega^2, 
\end{equation}
where Eqs.~(\ref{Kruskal_UV(r)}), (\ref{ModKS}) and (\ref{Second_conformal_BHtoWH}) are used. Meanwhile, $X_{\pm}=\sqrt{1-b/2M_{\pm}}e^{b/4M_{\pm}}$ is given by Eq.~(\ref{UV_mini_r}) and $k \equiv \frac{M_{+}}{M_{-}}\sqrt{f_{+}(b)/f_{-}(b)}$ is defined in Eq.~(\ref{Second_conformal_BHtoWH}). Notice that in this form, the factor causing singularity, $\left|\left(1-\frac{r_{+}}{2M_{+}}\right)e^{r_{+}/2M_{+}}\right|^{1-k}$, is without explicit dependence to $(V_{+}'', U_{+}'')$, which is similar to the physical RN solution given in Eq.~(\ref{RN_metric_in_outer}).

Starting from the coordinates of the resulting Penrose diagram, $(\tilde{V}, \tilde{U})$, 
 in which the metric is given by Eqs.~(\ref{Penrose_metric_lower}) and (\ref{Penrose_metric_upper}), the conformal transformation preserving the continuity at the thin shell can be represented as 
\begin{equation}\label{CT_of_the_final_Penrose_D}
\left(\bar{V}, \bar{U}\right)=\left(H(\tilde{V}), G(\tilde{U}) \right)
\end{equation}
where $(\bar{V}, \bar{U})$ are the coordinates of the other Penrose diagram as shown in Fig.~\ref{fig:CT_of_Penrose_D}. Notice that after this transformation, the trajectory of the thin shell, $r_{+}=r_{-}=b$, might not be a horizontal line anymore. Also, their construction is similar to that of the $(\tilde{V}, \tilde{U})$ coordinates, \textit{i.e.} the similar piling relation Eqs.~(\ref{BH_WH_connection_rule_0}) and (\ref{BH_WH_connection_rule}), and the inverse tangent function compactification. Thus, the non-compactified coordinates corresponding to $\left(\bar{V}, \bar{U}\right)$ should be given by
\begin{equation}
    (v_{\pm}, u_{\pm})=\left(\tan^{-1} \bar{V}_{\pm}, \tan^{-1} \bar{U}_{\pm} \right), 
\end{equation}
where $\left(\bar{V}_{-}, \bar{U}_{-}\right)=\left(\bar{V}, \bar{U}\right)$ and $\left(\bar{V}_{+}, \bar{U}_{+}\right)=\left(\bar{U}-\pi/2, \bar{V}-\pi/2 \right)$. Then the transformation Eq.~(\ref{CT_of_the_final_Penrose_D}) can be decomposed 
into 
\begin{equation}\label{CT_Penrose_lower}
(v_{-}, u_{-})=\left(h_{-}(V_{-}'), g_{-}(U_{-}') \right), 
\end{equation}
and 
\begin{equation}\label{CT_Penrose_upper}
(v_{+}, u_{+})=\left(h_{+}(V_{+}''), g_{+}(U_{+}'')  \right).  
\end{equation}

Now, we can discuss the conditions of the functions $h_{\pm}$ and $g_{\pm}$ defined in Eqs.~(\ref{CT_Penrose_lower}) and (\ref{CT_Penrose_upper}). Firstly, since they are transformations between the two Penrose diagrams shown in Fig.~\ref{fig:CT_of_Penrose_D},  the functions $h_{\pm}$ and $g_{\pm}$ must satisfy the following relations
\begin{equation}\label{odd_function_simp}
\begin{split}
h_{\pm}(0)& =g_{\pm}(0)=0   \\
h_{\pm}(x)& >0 \ and \ g_{\pm}(x)>0 \quad if \ x>0 \\
h_{\pm}(x)& <0 \ and \ g_{\pm}(x)<0 \quad if \ x<0. 
\end{split}
\end{equation}   
Moreover, since a null geodesic crossing the shell must have either one of the following two forms: $v_{+}=const.<0$ and $v_{-}=const.>0$ (for instance, the blue line in Fig.~\ref{fig:CT_of_Penrose_D}), or $u_{+}=const.<0$ and $u_{-}=const.>0$, without losing generality, we can simplify Eq.~(\ref{odd_function_simp}) into the condition that $h_{\pm}$ and $g_{\pm}$ are odd functions.     

Next, due to the fact that the continuity of the Penrose diagram at the shell must be preserved after the transformation, $h_{\pm}$ and $g_{\pm}$ are not independent. To begin with, we relate the two coordinates systems $\left(\bar{V}, \bar{U}\right)$ and $(\tilde{V}, \tilde{U})$ by using Eqs.~(\ref{CT_Penrose_lower}) and (\ref{CT_Penrose_upper}) as follows. For the lower region, we have 
\begin{equation}\label{Null_U_lower}
\bar{U}=\bar{U}_{-}=\tan^{-1}u_{-}=\tan^{-1}\left[g_{-}(U'_{-})\right]=\tan^{-1}\left[g_{-}\left(\tan(\tilde{U})\right)\right], 
\end{equation} 
and similarly
\begin{equation}\label{Null_V_lower}
\bar{V}=\tan^{-1}\left[h_{-}\left(\tan(\tilde{V})\right)\right]. 
\end{equation}
For the upper region, we have 
\begin{equation}\label{Null_U_upper}
\begin{split}
\bar{U}=\bar{U}_{+}+\frac{\pi}{2}=\tan^{-1}v_{+}+\frac{\pi}{2}&=\tan^{-1}\left[h_{+}(V''_{+})\right]+\frac{\pi}{2} \\
&=\tan^{-1}\left[h_{+}\left(\tan(\tilde{U}-\frac{\pi}{2})\right)\right]+\frac{\pi}{2} \\
&=-\tan^{-1}\left[h_{+}\left(\cot\tilde{U}\right)\right]+\frac{\pi}{2}, 
\end{split}
\end{equation} 
where the condition that $h_{+}$ and $\tan^{-1}$ are odd functions are used to obtain the last equality. Similarly, we have 
\begin{equation}\label{Null_V_upper}
\bar{V}= -\tan^{-1}\left[g_{+}\left(\cot\tilde{U}\right)\right]+\frac{\pi}{2}.  
\end{equation}
Then, the continuity of the null geodesics in the Penrose diagram equates Eq.~(\ref{Null_U_lower}) to Eq.~(\ref{Null_U_upper}), and Eq.~(\ref{Null_V_lower}) to Eq.~(\ref{Null_V_upper}). Then, from Eqs.~(\ref{Null_U_lower}) and (\ref{Null_U_upper}) we have
\begin{equation}\label{g_lower_as_f}
g_{-}(x)=\frac{1}{h_{+}(\frac{1}{x})}, 
\end{equation}
where the identity, $\tan^{-1} \left[x \right]+\tan^{-1}\left[1/x\right]=\pi/2$ when $x>0$,  is used. Similarly, from Eqs.~(\ref{Null_V_lower}) and (\ref{Null_V_upper}), we have 
\begin{equation}\label{f_lower_as_g}
h_{-}(x)=\frac{1}{g_{+}(\frac{1}{x})}. 
\end{equation}
Substituting Eqs.~(\ref{g_lower_as_f}) and (\ref{f_lower_as_g}) into Eq.~(\ref{CT_Penrose_lower}), we then have the relation that for a given transformation in the upper region, Eq.~(\ref{CT_Penrose_upper}), the transformation in the lower region must be of the form
\begin{equation}\label{CT_relation}
(v_{-}, u_{-})=\left(\frac{1}{g_{+}(1/V_{-}')}, \frac{1}{h_{+}(1/U_{-}')} \right).  
\end{equation}
With this relation, we can examine if there exist conformal transformations Eq.~(\ref{CT_of_the_final_Penrose_D}) able to fix the singularity of the metric at the event horizon in the upper region Eq.~(\ref{metric_after_2nd_trans}): 
\begin{equation}\label{eq37}
d s_{+}^2=-\frac{32M_{+}^3}{r_{+}}e^{-r_{+}/2M_{+}}\frac{X_{+}^2}{k^2} \left|V''_{+}U''_{+}\right|^{\frac{1-k}{k}} d V''_{+} d U''_{+}+r_{+}^2 d \Omega^2, 
\end{equation}
 while also able to keep the metric regular at the event horizon of the lower region, which in terms of $(V'_{-}, U'_{-})$ coordinates is given by 
\begin{equation}\label{lower_metric_in_VU'}
d s_{-}^2=-\frac{32M_{-}^3}{r_{-}}e^{-r_{-}/2M_{-}}X_{-}^2d V'_{-} d U'_{-}+r_{-}^2 d \Omega^2.
\end{equation}

By using Eqs.~(\ref{CT_Penrose_upper}) and (\ref{CT_relation}), we have the differential relations:
\begin{equation}\label{upper_dif_relation}
d V''_{+} d U''_{+}=\left(g'_{+}(U''_{+})h'_{+}(V''_{+})\right)^{-1} dv_{+}du_{+},
\end{equation}
and 
\begin{equation}\label{lower_dif_relation}
d V'_{-} d U'_{-}=
\frac{V'^2_{-}U'^2_{-}g^2_{+}\left(1/V'_{-}\right)h^2_{+}\left(1/U'_{-}\right)}{g'_{+}(1/V'_{-})h'_{+}(1/U'_{-})} dv_{-}du_{-},
\end{equation}
where the prime on the functions $g_{+}$ and $h_{+}$ represents taking derivative, \textit{i.e.} $g'_{+}(x)=dg_{+}/dx$ and $h'_{+}(x)=dh_{+}/dx$, while the primes on the rest of the letters are just symbols to distinguish different quantities. Substituting Eq.~(\ref{upper_dif_relation}) into the Eq.~(\ref{eq37}) and neglecting the irrelevant factors, we then have 
\begin{equation}\label{cancelation_upper_diff}
\left|V''_{+}U''_{+}\right|^{\frac{1-k}{k}}d V''_{+} d U''_{+}=\left|V''_{+}U''_{+}\right|^{\frac{1-k}{k}}\left(g'_{+}(U''_{+})h'_{+}(V''_{+})\right)^{-1} dv_{+}du_{+}.     
\end{equation}
Due to the symmetry and separability of $V''_{+}$ and $U''_{+}$ in the above relation, it is sufficient only to consider only the cases with $g'_{+}(x)=h'_{+}(x)$.  It is not difficult to see that in order to remove the singularity of the metric at the event horizons $V''_{+}=v_{+}=0$ or $U''_{+}=u_{+}=0$ in Eq.~(\ref{cancelation_upper_diff}), the function $h_{+}(x)$ and $g_{+}(x)$ must satisfy the following differential equation 
\begin{equation}\label{diff_eq_for_f}
h_{+}'(x)=g_{+}'(x)=\alpha(x)|x|^{\frac{1}{k}-1}, 
\end{equation}
where $\alpha(x)$ is a positive even function with $\alpha(0)$ is some finite and positive constant. Notice the requirement that $\alpha(x)$ to be a positive even function is to be consistent with the previous simplification that $h_{+}(x)$ is an odd function.

We first consider the simplest case, in which $\alpha(x)=\alpha_0$ is a positive constant. In this case, Eq.~(\ref{diff_eq_for_f}) has solution
\begin{equation}\label{f_plus_simplest_case}
h_{+}(x)=g_{+}(x)=sgn(x)\alpha_0 k |x|^{\frac{1}{k}}, 
\end{equation}
where $sgn(x)$ is the sign function, and the constant of integration is set to zero since $h_{+}(x)$ and $g_{+}(x)$ are odd functions. Substituting Eq.~(\ref{f_plus_simplest_case}) into Eq.~(\ref{lower_dif_relation}), 
we then have
\begin{equation}\label{lower_diff_simp_alpha}
d V'_{-} d U'_{-}=(\alpha_0)^2
k^2\left|U_{-}'V_{-}'\right|^{1-\frac{1}{k}} dv_{-}du_{-}.  
\end{equation}
By substituting Eq.~(\ref{lower_diff_simp_alpha}) into Eq.~(\ref{lower_metric_in_VU'}), we can see that the metric becomes singular at the event horizon $V'_{-}=v_{-}=0$ or $U'_{-}=u_{-}=0$. In fact, one can show that this result corresponds to the result of performing the second conformal transformation, Eq.~(\ref{Second_conformal_BHtoWH}), on the lower region instead of on the upper region to fix the discontinuity at the transition surface. 

Now, to show that no simple $r$-dependent  conformal coordinates can cover the event horizons in both the upper and the lower region at the same time, we first substitute Eqs.~(\ref{upper_dif_relation}) and (\ref{diff_eq_for_f}) into Eq.~(\ref{eq37}) to have
\begin{equation}\label{upper_metric_alpha}
d s_{+}^2=-\frac{32M_{+}^3}{r_{+}}e^{-r_{+}/2M_{+}}\frac{X_{+}^2}{k^2} \frac{1}{\alpha(V''_{+})\alpha(U''_{+})} d v_{+} d u_{+}+r_{+}^2 d \Omega^2, 
\end{equation}
and substitute Eqs.~(\ref{lower_dif_relation}) and (\ref{diff_eq_for_f}) into Eq.~(\ref{lower_metric_in_VU'}) to have
\begin{equation}\label{lower_metric_alpha}
d s_{-}^2=-\frac{32M_{-}^3}{r_{-}}e^{-r_{-}/2M_{-}}X_{-}^2 \frac{ \left|V'_{-}U'_{-}\right|^{\frac{1}{k}+1} h^2_{+}\left(1/V'_{-}\right)h^2_{+}\left(1/U'_{-}\right)}{\alpha(1/V'_{-})\alpha(1/U'_{-})} dv_{-}du_{-} +r_{-}^2 d \Omega^2,
\end{equation}
where $h_{+}(x)=g_{+}(x)$ is used. Although we do not have an explicit relation between $\alpha(x)$ and $h_{+}(x)$ since the function $\alpha(x)$ is not determined, we still can see several features of the metric from Eqs.~(\ref{upper_metric_alpha}) and (\ref{lower_metric_alpha}).  Firstly, it is worth mentioning that one special feature of Eq.~(\ref{lower_metric_alpha}) is that the behavior of the metric component at the event horizons $V'_{-}=v_{-}=0$ or $U'_{-}=u_{-}=0$ is affected by the behavior of $\alpha(x)$ and $h_{+}(x)$ approaching infinity, i.e. $\lim_{x\to\infty} \alpha(x)$ and $\lim_{x\to\infty} h_{+}(x)$.   Since the function $\alpha(x)$ enters Eq.~(\ref{upper_metric_alpha}) directly, then we can say that if the continuity of the metric is preserved, the behavior of the metric at the event horizon in the lower region is related to that at the infinity in the upper region and vice versa. One also can show that if $\lim_{x\to\infty} \alpha(x)$ is finite and positive, Eq.~(\ref{lower_metric_alpha}) is always singular at the event horizons.

Next, from Eq.~(\ref{upper_metric_alpha}), we can see that to have the metric components be simple $r$-dependent, the function $\alpha(x)$ must satisfy the relation 
\begin{equation}\label{non_static_proof}
\alpha(V''_{+})\alpha(U''_{+})=\gamma(V''_{+}U''_{+}),    
\end{equation}
since $V''_{+}U''_{+}$ only depends on $r_{+}$. Now, one can show that for $\alpha(x)$ and $\gamma(x)$ to be differentiable functions and satisfy the above relation (\ref{non_static_proof}), up to a constant, they must be of the form 
\begin{equation}\label{form_alpha}
\alpha(x)=\gamma(x)=x^a,   
\end{equation}
with $a$ is a real number. Notice that Eq.~(\ref{non_static_proof}) can be relaxed to the relation $\alpha(x)\beta(y)=\gamma(xy)$, and the result (\ref{form_alpha}) still holds, \textit{i.e.} $\alpha(x)=\beta(x)=\gamma(x)=x^a$ up to the trivial constants. The proof of this statement is given as follows: \\ 
If $\alpha(x)$, $\beta(y)$ and $\gamma(xy)$ are not some trivial constant functions, by taking derivative with respect to $x$ and $y$, we have 
\begin{equation*}
\begin{split}
     \alpha'(x)\beta(y)&=y\gamma'(xy)  \\
        \alpha(x)\beta'(y)&=x\gamma'(xy),
\end{split}
\end{equation*} 
respectively. 
Dividing the above two equations, we have 
\begin{equation*}
x\frac{\alpha'(x)}{\alpha(x)}=y\frac{\beta'(y)}{\beta(y)}=a,
\end{equation*}
with $a$ is some non-zero constant. Thus, we have solutions for $\alpha(x)$ and $\beta(y)$ as 
\begin{equation*}
\begin{split}
\ln\alpha(x)&=a\ln x+C_1 \\
\ln\beta(x)&=a\ln x+C_2, 
\end{split}
\end{equation*}
which can be rewritten as $\alpha(x)=C_1x^a$ and $\beta(x)=C_2x^a$. With the relation  $\alpha(x)\beta(y)=\gamma(xy)$, we then have $\gamma(x)=C_1C_2x^a$. Thus, up to the trivial constants, $C_1$ and $C_2$, we have  $\alpha(x)=\beta(x)=\gamma(x)=x^a$. Q.E.D.

However, the only possible choice of Eq.~(\ref{form_alpha}) making Eq.~(\ref{upper_metric_alpha}) regular at the event horizon is $a=0$, which is nothing but the simplest choice, $\alpha(x)=\alpha_0$, we discussed earlier. Since we have shown that fixing metric in the upper region by choosing $\alpha(x)=\alpha_0$ creates singularity at the event horizons in the lower region through Eq.~(\ref{lower_diff_simp_alpha}); therefore, no simple $r$-dependent global conformal coordinates can cover both the upper and lower regions without having any singularity at event horizons. 
      
Next, we try an ansatz $\alpha(x)=\left|x\right|^{\gamma}+C$ where $\gamma$ is real and $C$ is real and positive. Substituting this ansatz into Eq.~(\ref{diff_eq_for_f}), we have
\begin{equation}\label{diff_f_ansatz}
h'_{+}(x)=\left|x\right|^{\gamma+\frac{1}{k}-1}+C\left|x\right|^{\frac{1}{k}-1}, 
\end{equation}
which has solution 
\begin{equation}\label{ansatz_solution}
h_{+}(x)=\left(\frac{1}{\gamma+\frac{1}{k}}\left|x\right|^{\gamma+\frac{1}{k}}+C k \left|x\right|^{\frac{1}{k}}\right)sgn(x). 
\end{equation}
Then the relevant factor in Eq.~(\ref{lower_dif_relation}) at the event horizon is of the form 
\begin{equation}\label{the_relevant_factor_ansatz}
\lim_{x\to 0} \frac{x^2h^2_{+}(1/x)}{h'_{+}(1/x)}=\lim_{x\to 0} x^2 \left(\frac{1}{\gamma+\frac{1}{k}}\left|\frac{1}{x}\right|^{\gamma+\frac{1}{k}}\right)^2 \left|x\right|^{\gamma+\frac{1}{k}-1} = \lim_{x\to 0}  \frac{1}{\left(\gamma+\frac{1}{k}\right)^2} \left|x\right|^{-\gamma-\frac{1}{k}+1}, 
\end{equation}
if $\gamma>0$. It is then easy to see that to have Eq.~(\ref{the_relevant_factor_ansatz}) to be a finite value, we must have $\gamma=1-1/k>0$, which is only possible when $k>1$ since $k \equiv \frac{M_{+}}{M_{-}}\sqrt{f_{+}(b)/f_{-}(b)}>0$. Substituting $\gamma=1-1/k$ into Eq.~(\ref{ansatz_solution}), we obtain the conformal transformation, which works only when $k>1$, fixing the singularity at the upper event horizons without causing singularity at the lower horizons as
\begin{equation}
h_{+}(x)=g_{+}(x)=\left(\left|x\right|+C k \left|x\right|^{\frac{1}{k}}\right)sgn(x),  
\end{equation}
with the corresponding $\alpha(x)=\left|x\right|^{1-\frac{1}{k}}+C$ and $f'_{+}(x)=g'_{+}(x)=1+C\left|x\right|^{\frac{1}{k}-1}$. 
After this transformation, the metric of the upper region, Eq.~(\ref{upper_metric_alpha}), becomes
\begin{equation}\label{upper_metric_ansatz}
d s_{+}^2=-\frac{32M_{+}^3}{r_{+}}e^{-r_{+}/2M_{+}}\frac{X_{+}^2}{k^2} \frac{1}{\left(\left|V''_{+}\right|^{1-1/k}+C\right)\left(\left|U''_{+}\right|^{1-1/k}+C\right)} d v_{+} d u_{+}+r_{+}^2 d \Omega^2, 
\end{equation}
which is regular at either the event horizons $V''_{+}=v_{+}=0$ or $U''_{+}=u_{+}=0$. 
Meanwhile, the metric of the lower region, Eq.~(\ref{lower_metric_alpha}), reduces to 
\begin{equation}\label{lower_metric_ansatz}
d s_{-}^2=-\frac{32M_{-}^3}{r_{-}}e^{-r_{-}/2M_{-}}X_{-}^2  \frac{\left(\left|\frac{1}{V'_{-}}\right|^{1-\frac{1}{k}}+2Ck+C^2k^2\left|\frac{1}{V'_{-}}\right|^{\frac{1}{k}-1}\right)\left(\left|\frac{1}{U'_{-}}\right|^{1-\frac{1}{k}}+2Ck+C^2k^2\left|\frac{1}{U'_{-}}\right|^{\frac{1}{k}-1}\right)}{\left(\left|\frac{1}{V'_{-}}\right|^{1-1/k}+C\right)\left(\left|\frac{1}{U'_{-}}\right|^{1-1/k}+C\right)} dv_{-}du_{-} +r_{-}^2 d \Omega^2. 
\end{equation}
Now, although the function for the metric component given in Eq.~(\ref{lower_metric_ansatz}), strictly speaking, does not have value when $V'_{-}=v_{-}=0$ or $U'_{-}=u_{-}=0$, the metric at the event horizons can be defined by taking the limit of $V'_{-}=v_{-} \to 0$ or $U'_{-}=u_{-} \to 0$.
For instance, at the event horizon with $V'_{-}=v_{-}=0$, Eq.~(\ref{lower_metric_ansatz}) approaches to
\begin{equation}\label{lower_metric_ansatz_V=0}
d s_{-}^2=-\frac{32M_{-}^3}{r_{-}}e^{-r_{-}/2M_{-}}X_{-}^2  \frac{\left(\left|\frac{1}{U'_{-}}\right|^{1-\frac{1}{k}}+2Ck+C^2k^2\left|\frac{1}{U'_{-}}\right|^{\frac{1}{k}-1}\right)}{\left(\left|\frac{1}{U'_{-}}\right|^{1-1/k}+C\right)} dv_{-}du_{-} +r_{-}^2 d \Omega^2,
\end{equation}
where $U'_{-}$ is finite but not zero. 

Since in our two examples, we can always choose the side of spacetime with $k>1$ to perform the second transformation, the ansatz $\alpha(x)=\left|x\right|^{1-\frac{1}{k}}+C$, with $C >0$, works providing that the metric at the event horizons is defined by taking the limit as the form given in Eq.~(\ref{lower_metric_ansatz_V=0}). Also, notice that $V''_{+}=v_{+}=0$ and $U''_{+}=u_{+}=0$ together include both the white hole and black hole event horizons in the upper region. The same goes for $V'_{-}=v_{-}=0$ or $U'_{-}=u_{-}=0$ in the lower region. We then have regular metric solutions at all event horizons appearing in Fig.~\ref{fig:CT_of_Penrose_D}.  

Next, we comment on the metric solutions Eqs.~(\ref{upper_metric_ansatz}) and (\ref{lower_metric_ansatz}) on spacetime points not at event horizons. We first consider the region with a finite value of $r_{\pm}$ other than $2M_{\pm}$. Since  $k>0$ by definition and $C>0$ for the ansatz $\alpha(x)=\left|x\right|^{\gamma}+C$ we use, these two conditions guarantee that the factors inside the parentheses shown in Eqs.~(\ref{upper_metric_ansatz}) and (\ref{lower_metric_ansatz}) are also finite and nonzero when $(U''_{+}, V''_{+})$ and $(U'_{-}, V'_{-})$ are finite and nonzero. Therefore, Eqs.~(\ref{upper_metric_ansatz}) and (\ref{lower_metric_ansatz}) are regular when $r_{\pm} \neq 2M_{\pm}$ but finite. For the asymptotic regime, $r_{\pm} \to \infty$, the metric components $g_{v_{+}, u_{+}}$ and $g_{v_{-}, u_{-}}$ approach zero in Eqs.~(\ref{upper_metric_ansatz}) and (\ref{lower_metric_ansatz}) respectively. However, this behavior happens only at infinity, which is a standard feature for the Kruskal-like solution for the Schwarzschild solution Eq.~(\ref{Kruskal_in_U_V}).

\begin{figure}[h!]
	\centering	
         \includegraphics[scale=0.5]{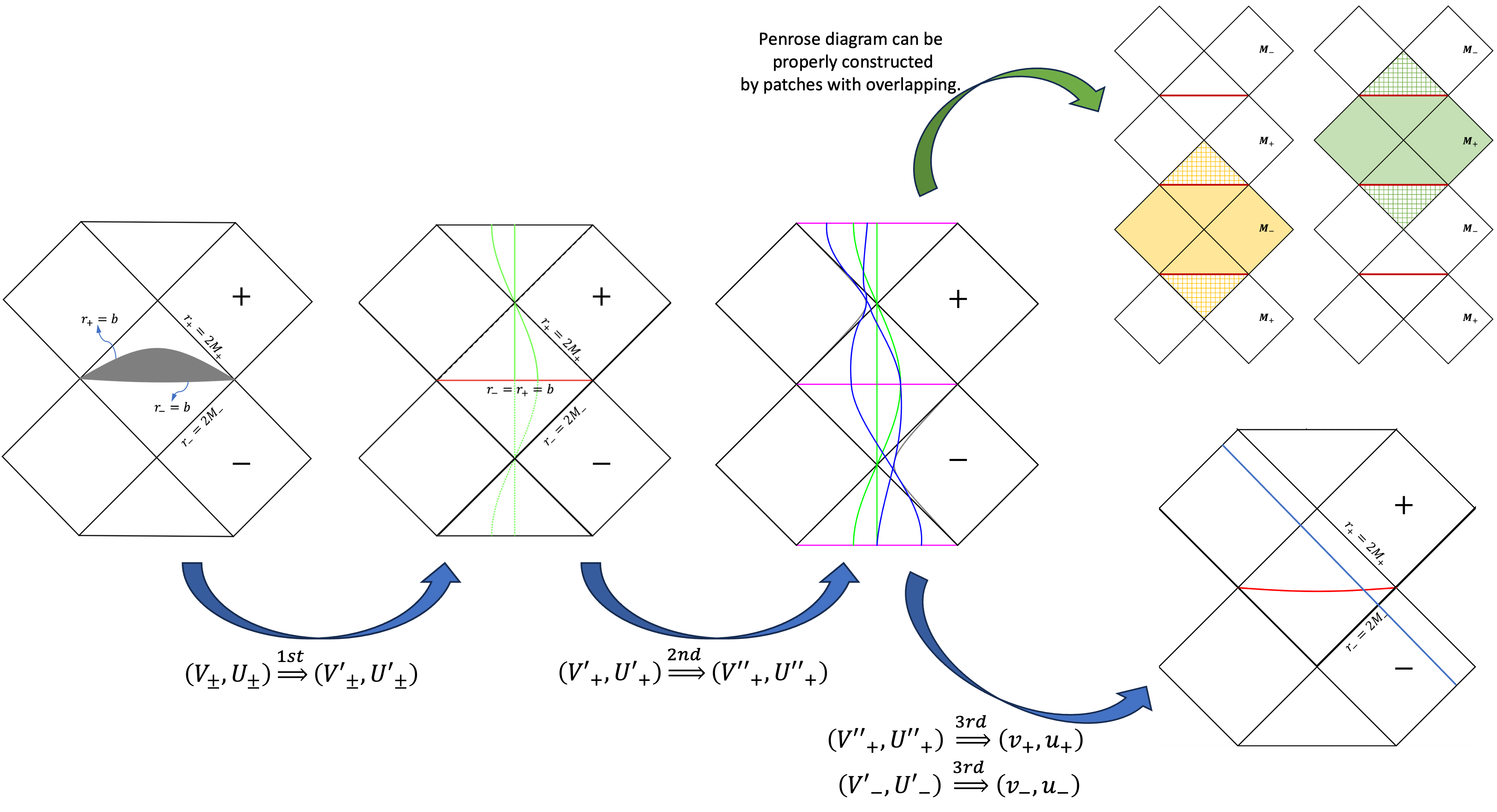}
	\caption{ A cartoon picture summarizes the three conformal transformations used to construct the global conformal coordinates of the generalized black-to-white hole bounce model with $M_{+}>M_{-}$. 
 These transformations are given by Eqs.~(\ref{ModKS}), (\ref{Second_conformal_BHtoWH}), and (\ref{3_trans_summary}), respectively. Notice that although the figures are all represented by using the compactified conformal coordinates, the transformations are given in terms of the non-compactified conformal coordinates. Also, the Penrose diagram for such a model can be properly constructed without introducing the third transformation as discussed in Sec.~\ref{Subsec:RN}. }  	\label{fig:Summary}
\end{figure}

We end this subsection first with a recap of the third transformation and a cartoon picture summarizing the transformations studied in this work. We then comment on the types of cut-and-pasted spacetimes for which these transformations give a global conformal coordinate chart.

The third transformation used to remove the coordinate singularity at the event horizon (on the side subjected to the second transformation) while preserving the continuity of the coordinates at the shell is given by  
\begin{equation}\label{3_trans_summary}
    \begin{split}
    (v_{+}, u_{+})&=\left(h_{+}(V_{+}''), g_{+}(U_{+}'')  \right), \\
        (v_{-}, u_{-})&=\left(\frac{1}{g_{+}(1/V_{-}')}, \frac{1}{h_{+}(1/U_{-}')} \right),  \\
        with \quad h_{+}(x)&=g_{+}(x)=\left(\left|x\right|+C k \left|x\right|^{\frac{1}{k}}\right)sgn(x), 
    \end{split}
\end{equation}
where $M_{+}>M_{-}$ is assumed (so $k>1$). The procedures of all transformations studied related to this case are summarized in Fig.~\ref{fig:Summary}. If $M_{+}<M_{-}$, the second transformation should apply to the lower region so a similar third transformation can be applied to solve the coordinate singularity caused by the second transformation. That is, the third transformation we designed can only apply to the stretched type of singularity. Although here we only use the generalized black-to-white hole bounce to discuss the formalism of the third transformation preserving the continuity of the coordinates at the shell, the method can be applied similarly to the Schwarzschild-to-de Sitter model.     
 
\begin{figure}[h!]
	\centering	
         \includegraphics[scale=0.3]{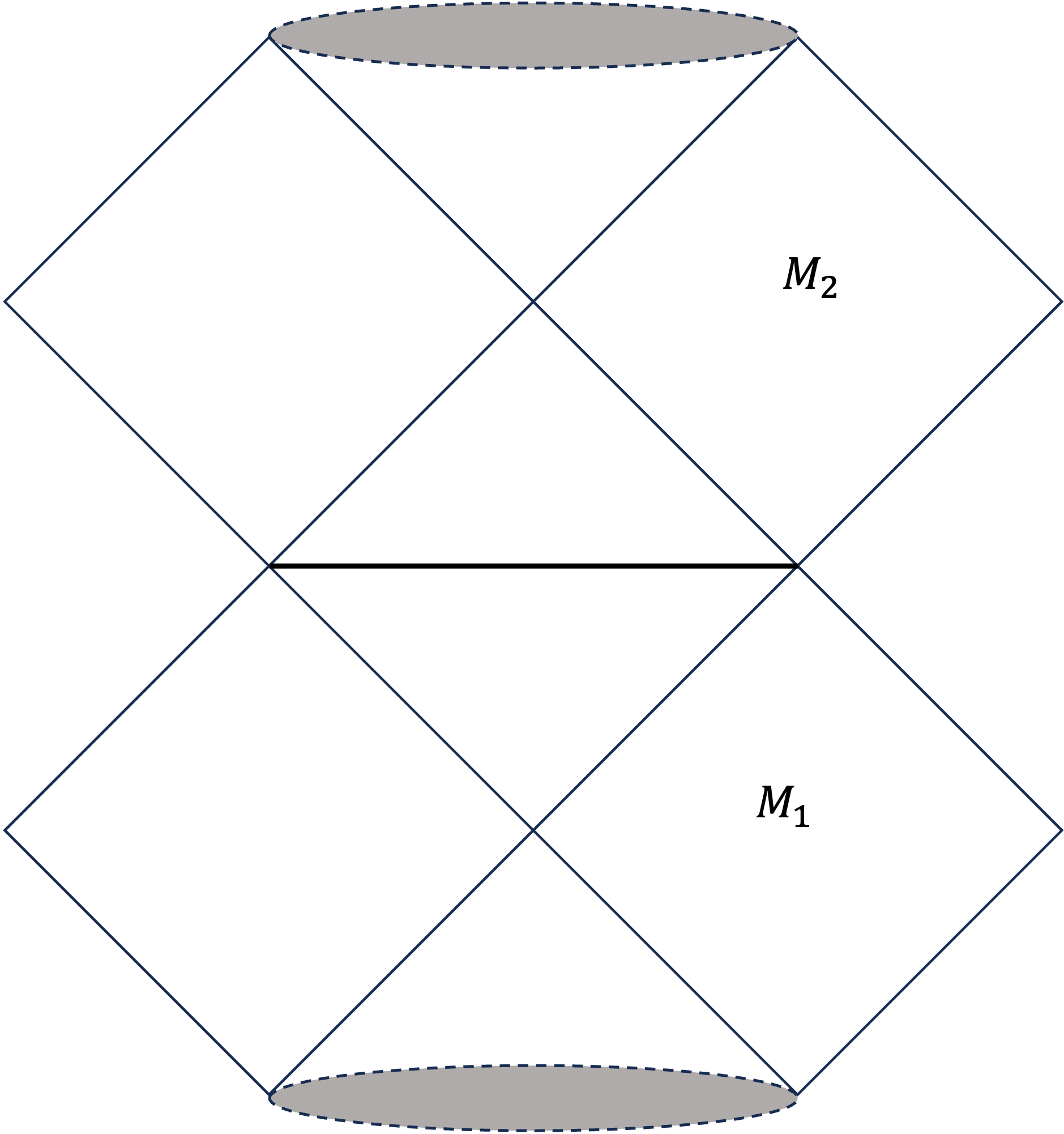}
	\caption{In a more complicated model where the gray areas are connected to some arbitrary manifolds, the procedure introduced in this work can generate a ``local'' chart covering the entire white region providing that the shell is static.}  	\label{fig:local}
\end{figure}

Lastly, it is also important to note that these transformations can result in a global conformal coordinate chart only when the spacetime is cyclic with a fixed transition radius. For instance, in the generalized black-to-white hole bouncing model, these transformations only apply to a cyclic model between $M_{-}$ and $M_{+}$ at the same transition radius $b$, see Fig.~\ref{fig:tiling}. This can be understood by observing that the second and third transformations described here are specified by the parameter $k \equiv \frac{M_{+}}{M_{-}}\sqrt{f_{+}(b)/f_{-}(b)}$. Nevertheless, in a more complicated case, the procedure introduced here offers a ``local'' chart covering the entire static shell and the adjacent event horizons on each side, see Fig.~\ref{fig:local}.

\section{Conclusions and outlooks }\label{Sec:conclusion}

 Using two examples, the generalized black-to-white hole transition and the Schwarzschild-de Sitter transition, we study the coordinates of Penrose diagrams for models constructed through a static spacelike thin shell. We discuss the required coordinate transformations to obtain the conformal coordinates of the Penrose diagrams for these two spacetimes and point out that an implicit discontinuity exists after the coordinate transformation corresponding to a simple cut-and-paste. We argue that this issue is general for models of this type. To fix this issue while keeping null geodesics at $45\degree$ in the resulting diagram, we introduce the second transformations (\ref{Second_conformal_BHtoWH}) and (\ref{Second_conformal_BH_dS}) for the above mentioned two models respectively. However, this second transformation unavoidably reintroduces a different type of coordinate singularity back to the event horizon, which causes the degeneracy between four-vectors at the event horizon. Thus, the resulting conformal coordinates after the second transformation still cannot be a global coordinate for the spacetime models considered in this work. Nevertheless, they can serve as a continuous conformal coordinate system covering a part of spacetime containing the entire shell. In this coordinate system, the spacetime metric components and the trajectories of geodesics are continuous at the shell. With the second transformation, the Penrose diagram can be constructed properly since different patches overlap.

 In Sec.~\ref{Subsec_3rdTrans}, we further demonstrate how to construct the third conformal transformation, which preserves the continuity of the coordinates at the shell. Using this formalism, we provide a transformation suitable for solving the stretching type of coordinate singularity,  \textit{i.e.} when $k>1$ and $k_{dS}>1$. In the resulting coordinate system, the metric components Eqs.~(\ref{upper_metric_ansatz}) and (\ref{lower_metric_ansatz}) are regular everywhere with the condition that the metric components at the event horizon must be defined by taking the limit, Eq.~(\ref{lower_metric_ansatz_V=0}). A 
 cartoon picture summarizing the three transformations is given in Fig.~\ref{fig:Summary}.
 
 On the other hand, the third transformation found in this work cannot be applied to the squeezing type of coordinate singularity, \textit{i.e.} when $k<1$ and $k_{dS}<1$. Although less well-known, the squeezing type of coordinate singularity also exists when one tries to cover the inner horizon of the physical Reissner-Nordstr{\"o}m (RN) solution Eq.~(\ref{RN_metric_in_outer}) by using the Kruskal-like coordinate chart constructed for the outer horizon and vice versa. Then, it is an interesting question if the squeezing type of coordinate singularity can be removed by some clever choice of ansatz $\alpha(x)$, which is related to the third transformation by the differential equation (\ref{diff_eq_for_f}).
 Then, such a coordinate transformation should also be able to provide the global conformal coordinates for the RN solution.
 However, the study of this question is beyond the scope of this work, and we would like to put it into future works.

\newpage

\section*{Data availability}
The data-sets generated or analyzed during this study are included in this article.

\section*{Acknowledgments}
W.L. would like to thank Ting-Wei Tsui for several useful discussions. W.L. is supported by the National Research Foundation of Korea (NRF) grant funded by the Korea government (MSIT) (2021R1A4A5031460). DS is partially supported by the US National Science Foundation, under Grants No.
PHY-2014021 and PHY-2310363. DY is supported by the National Research Foundation of Korea (Grant No.: 2021R1C1C1008622, 2021R1A4A5031460).



\end{document}